%% file: main.tex
\documentclass[aps,showpacs,twocolumn,prd,superscriptaddress,nofootinbib,floats,floatfix]{revtex4-1}

\usepackage[T1]{fontenc}
\usepackage[utf8]{inputenc}
\usepackage{graphicx}
\usepackage{xspace}
\usepackage{amsmath}
\usepackage{amsfonts}
\usepackage{amssymb}
\usepackage{latexsym}
\usepackage{microtype}
\usepackage{enumerate}
\usepackage{tabularx}
\usepackage{lmodern}
\usepackage{verbatim}
\usepackage{soul}
\usepackage[dvipsnames,x11names,svgnames,rgb,table]{xcolor}
\usepackage[colorlinks]{hyperref}
\usepackage{orcidlink}

\DeclareMathAlphabet{\mathpzc}{OT1}{pzc}{m}{it}

\definecolor{amethyst}{HTML}{a45ee5}
\definecolor{austriawien}{HTML}{441678}
\definecolor{dodgerblue}{HTML}{1E90FF}
\definecolor{deepgreen}{HTML}{009141}

\colorlet{papercolor}{amethyst}

\hypersetup{
     colorlinks=true,
     linkcolor=papercolor,
     filecolor=papercolor,
     citecolor = papercolor,      
     urlcolor=papercolor,
}

\newcommand{\pd}{\partial}

\newcommand{\AKerr}{A^{\rm Kerr}}
\newcommand{\Bnp}{B^{\rm Kerr}_{np}}
\newcommand{\Bnpa}{B^{\rm Kerr}_{npa}}

\newcommand{\bfa}{\boldsymbol{a}}

\newcommand{\bfL}{\boldsymbol{L}}

\newcommand{\pphi}{p_{\phi}}

\newcommand{\bham}{\affiliation{School of Physics and Astronomy, University of Birmingham, Edgbaston, Birmingham, B15 2TT, United Kingdom}}

\newcommand{\igwa}{\affiliation{Institute for Gravitational Wave Astronomy, University of Birmingham, Edgbaston, Birmingham, B15 2TT, United Kingdom}}

\begin{document}

\title{Strong Field Scattering of Two Black Holes: Exploring Gauge Flexibility}

\author{Adam Clark \orcidlink{0009-0001-0551-2481}}
\email{axc157@student.bham.ac.uk}
\bham 
\igwa

\author{Geraint Pratten \orcidlink{0000-0003-4984-0775}}
\email{g.pratten@bham.ac.uk}
\bham 
\igwa

\hypersetup{pdfauthor={Pratten et al.}}
\date{\today}

\begin{abstract}
Recent advances in post-Minkowskian (PM) gravity provide new avenues for the high precision modeling of compact binaries. In conjunction with the effective one body (EOB) formalism, highly accurate PM informed models of binary black holes on scattering trajectories have emerged.
Several complementary approaches currently exist, in particular the SEOB-PM model, the $w_{\rm eob}$ framework and the recent Lagrange-EOB (LEOB) approach. 
These models incorporate PM results in fundamentally different ways, employing distinct resummation schemes and gauge choices. 
Notably, both SEOB-PM and LEOB have been used to compute gravitational waves of bound systems, showing excellent agreement with numerical relativity (NR).
The essential component to all of the models is the EOB mass-shell condition describing the dynamics of the two-body spacetime. 
In this work we will investigate how this mass-shell condition is constructed, paying particular attention to the impact of gauge choices and how they interact with different resummation schemes, showing that there is a strong dependence on both coordinate choice and EOB gauge.
For the region of parameter space considered, we find that the best performing gauges coincide with the choices made in SEOB-PM and $w_{eob}$, with other choices exhibiting worse performance. 
The case of spinning black holes is also considered, where the current techniques for spinning EOB-PM are reviewed and compared. 
We also introduce a new gauge based upon the centrifugal radius, which improves upon previous models, particularly for large and negative spins. 
This offers a promising avenue for further resummation of spin information within the EOB-PM framework.
\end{abstract}

\maketitle

\section{Introduction}
Gravitational wave observations of compact binaries provide a unique and powerful window into the physics of strong gravitational fields. The ever growing catalog of binary black holes (BBH), binary neutron stars (BNS), and neutron star-black hole binaries (NSBH) \cite{LIGOScientific:2018mvr, LIGOScientific:2020ibl, KAGRA:2021vkt} provides an excellent opportunity to test fundamental physics. At the core of this problem is the need to accurately model the dynamics and emitted gravitational waves of two-body systems within general relativity.

Several analytical approximation schemes have been developed to address this problem, each suited to different regions of the parameter space.
One of the most widely used is the post-Newtonian (PN) expansion~\cite{Blanchet:2013haa}, which expands in weak fields and small velocities $[\epsilon \sim GM/(rc^2) \sim (v/c) \ll 1]$.
Such PN results have been extensively incorporated into the semianalytical effective-one-body (EOB) formalism~\cite{Buonanno:1998gg,Buonanno:2000ef,Nagar:2018zoe,Chiaramello:2020ehz,Pompili:2023tna,Ramos-Buades:2023ehm,Khalil:2023kep,Gamboa:2024imd,Nagar:2024dzj}, with calibration against numerical relativity (NR) simulations playing a key role in completing the models through merger and ringdown.
An alternative approach is the post-Minkowskian (PM) expansion \cite{Bel:1981be, Westpfahl:1979gu, Westpfahl:1985tsl, Blanchet:1985sp}, which is a weak-field expansion $[\epsilon \sim GM/(rc^2) \ll 1]$ that retains all orders in velocity, effectively resumming the PN expansion.
The PM expansion has seen rapid growth in recent years due to the novel connections between classical scattering and quantum scattering amplitudes identified in~\cite{Damour:2016gwp, Damour:2017zjx, Damour:2019lcq}.
Key results have come from a variety of formalisms including scattering amplitudes, \cite{Bern:2019nnu, Bern:2020buy, Bern:2021dqo, Bern:2021yeh}, worldline effective field theory (WEFT) \cite{Kalin:2020mvi, Kalin:2022hph, Kalin:2020fhe, Dlapa:2021npj, Dlapa:2021vgp, Dlapa:2022lmu}, and worldline quantum field theory (WQFT) \cite{Mogull:2020sak, Jakobsen:2021smu, Jakobsen:2022psy}. 
The current state of the art for nonspinning scattering is the recent radiative 5PM calculation, which includes information up to the first order in self-force (5PM-1SF) \cite{Driesse:2024xad, Driesse:2024feo}. 

In the spinning sector, both classical and field-theoretic methods have been used to compute a variety of spin corrections. 
At $1$PM, it has been shown that the test-mass limit entirely determines the scattering angle to all orders in spin \cite{Vines:2017hyw}. 
Such all order results have also been computed via scattering amplitudes at 1PM and 2PM \cite{Aoude:2022thd, Aoude:2023vdk}. 
At higher orders, quartic-in-spin corrections at 3PM were recently computed in \cite{Akpinar:2025bkt} for a single spinning black hole and spin-orbit corrections also available up to 4PM \cite{Jakobsen:2023hig}.

Given the impressive progress in perturbative calculations, it is of great interest to the waveform modeling community to incorporate PM results into models suitable for gravitational wave data analysis, such as EOB~\cite{Buonanno:1998gg,Buonanno:2000ef}. 
Since the PM expansion effectively resums the PN expansion at each order in $G$\footnote{The $(n+1)$PM order fully contains $n$PN plus an infinite series of velocity corrections}, it could prove invaluable when dealing with physical systems beyond the quasicircular limit, such as binaries on eccentric orbits or hyperbolic encounters, e.g.~\cite{Gupte:2024jfe,Morras:2025nlp,Morras:2025xfu,Planas:2025plq}. 
Transcribing PM information into bound-orbit models has been tackled in a variety of ways, including the direct calculation of PM Hamiltonians using EFT techniques \cite{Cheung:2018wkq} and the EOB formalism~\cite{Buonanno:2024byg,Damour:2025uka}. 
An alternative strategy is provided by the boundary-to-bound maps \cite{Kalin:2019rwq, Kalin:2019inp}, which relate gauge-invariant scattering observables to their bound counterparts, thus bypassing the need to work with gauge-dependent quantities.

However, there are still open and important challenges with these approaches. 
For example, in the case of misaligned spins, the mapping from scattering to bound observables is not yet fully understood.  
Moreover, it is well established that the PM expansion alone does not demonstrate good agreement against NR~\cite{Rettegno:2023ghr,Swain:2024ngs} and requires the use of resummation techniques to further improve the convergence of the results~\cite{Damour:2022ybd,Rettegno:2023ghr,Buonanno:2024vkx,Swain:2024ngs}.
In a series of works on gravitational scattering and the EOB formalism~\cite{Damour:2016gwp, Damour:2017zjx, Damour:2019lcq}, a systematic method was developed to incorporate PM information into the EOB framework through the gauge-invariant scattering angle. 
This methodology has enabled the construction of PM-based EOB models, notably the SEOB-PM model \cite{Buonanno:2024byg, Buonanno:2024vkx} and the $w_{\rm eob}$ framework \cite{Damour:2022ybd, Rettegno:2023ghr}. 
These models exhibit distinct behavior based on differing gauge choices and approaches toward resummation~\cite{Swain:2024ngs}. 
Exploring a range of possible gauge choices and the implications for the accuracy of the model will be the primary focus of this work.

Gauge flexibility has been explored in the literature to varying degrees. 
On the PN side, numerous results have been derived within the DJS gauge~\cite{Damour:2008qf}, including a recent systematic comparison of spin gauges presented in~\cite{Placidi:2024yld}. 
Within the EOB-PM models, recent work has developed a new Lagrange-multiplier based approach (LEOB)~\cite{Damour:2025uka}, including a discussion on a range of possible gauge choices, though with a strong focus on the Lagrange-Just-Boyer-Lindquist (LJBL) gauge employed in that work. 
A study on gauge choices in the context of the conservative dynamics of bound-orbit Hamiltonians up to 4PM was presented in~\cite{Khalil:2022ylj}, though limited to a single coordinate gauge, as was the treatment in \cite{Buonanno:2024vkx}. 
A systematic study of different gauge choices and the impact on the scattering angle in the strong-field regime has not yet been carried out. 
This is the gap that we aim to address in this work.

The paper will be structured as follows: Section \ref{section:scattering} will introduce the key ideas of EOB in the post-Minkowskian expansion. Section \ref{section:Non-Spinning} will construct a variety of EOB-PM models in the nonspinning sector and assess the impact of different gauge choices. Spin will be introduced in Section \ref{section:Spinning-Sector}, where we develop on previous works \cite{Buonanno:2024vkx, Khalil:2022ylj, Rettegno:2023ghr} and introduce an alternative approach to spinning EOB-PM.

Throughout this work we will employ geometric units, $c = G = 1$, though we will often explicitly track factors of $G$ to count the PM order. 
Component masses will be denoted by $m_1$ and $m_2$ respectively, the total mass by $M = m_1 + m_2$, the reduced mass by $\mu = m_1 m_2/M$, and the symmetric mass ratio $\nu = \mu/M$. 
We will often find it convenient to work with rescaled coordinates and momenta, defined throughout to be
\begin{align}
r &= \frac{R}{GM}, & t &= \frac{T}{GM}, \\
p_{\mu} &= \frac{P_{\mu}}{\mu}, & \pphi &= \frac{J}{G \mu M}.
\end{align}

\section{Gravitational Scattering and the Effective One Body Formalism} \label{section:scattering}

\subsection{The Kerr metric} 
\label{subsection:kerr-metric}
We begin by outlining different approaches to constructing a Kerr Hamiltonian for test particle dynamics, which provides the foundation for the spinning EOB Hamiltonians discussed later in this work. 
To that end, our starting point is the Kerr metric written in Boyer-Lindquist coordinates $(T, R, \theta, \varphi)$, given by~\cite{Wald:1984rg}
\begin{align} \label{Kerr-metric}
    g^{\mu\nu}_{\rm Kerr}\pd_\mu \pd_\mu &= -\frac{\Lambda}{\Delta \Sigma} \pd_T^2 + \frac{\Delta}{\Sigma}\pd_R^2 + \frac{1}{\Sigma}\pd_{\theta}^2 \\
    &\hspace{2em} + \frac{\Sigma - 2MR}{\Sigma \Delta \sin^2{\theta}}\pd_{\varphi}^2 - \frac{4MRa}{\Sigma\Delta}\pd_T\pd_\varphi, \notag
\end{align}
where $M$ is the black hole mass and $a$ is the Kerr parameter. The spin angular momentum is related to the Kerr parameter through $S = M a = G M^2 \chi$, where $S$ is the dimensionful spin angular momentum, $a$ is a mass-rescaled spin parameter (with dimensions of length) and $\chi$ is the dimensionless spin parameter. We further define the following useful combinations,
\begin{align}
    \Sigma &= R^2 + a^2\cos^2{\theta} \notag, \\
    \Delta &= R^2 - 2 M R + a^2 \notag, \\
    \Lambda &= (R^2 + a^2)^2 - a^2 \Delta \sin^2{\theta}. \notag 
\end{align}
The Kerr metric potentials can then be defined as follows~\cite{Khalil:2023kep}
\begin{align}
    \AKerr &= \frac{\Sigma \Delta}{\Lambda}, \\
    \Bnp &= \frac{R^2}{\Sigma}\left( \frac{\Delta}{R^2} - 1 \right), \\
    \Bnpa &= -\frac{R^2}{\Sigma \Delta}(\Sigma + 2MR).
\end{align}
\newline
The dynamics of a nonspinning timelike test particle of mass $\mu$ in the Kerr metric is obtained by solving the relativistic mass-shell condition~\cite{Buonanno:1998gg,Damour:2001tu}
\begin{align}\label{mass-shell-condition}
    g^{\mu\nu}_{\rm Kerr} \, P_{\mu}P_{\nu} + \mu^2 = 0,
\end{align}
where $P_{\mu} = (-P_0, P_R, P_{\theta}, P_{\varphi})$ and $P_0 = E_{\rm Kerr} = H_{\rm Kerr}$. 
In this work, we specialize to equatorial orbits, eliminating the dependence on $P_{\theta}$, and restricting the validity of our discussion to aligned-spin binaries. 
We will leave the case of spin precession to a future study. 
We can solve this condition for the Hamiltonian ${H}_{\rm Kerr} = H^{\rm Kerr}_{\rm orb} + H^{\rm Kerr}_{\rm SO}$, finding 
\begin{align} 
\label{Kerr-hamiltonian}
    & {H}_{\rm Kerr}  =  \frac{2 M P_{\varphi} a}{R^3 + a^2 (R + 2M)} \\
                             & \notag + \sqrt{\AKerr \left( \mu^2 + \frac{P_{\varphi}^2}{R^2} + (1 + \Bnp)P_{R}^2 + \Bnpa\frac{P_{\varphi}^2 a^2}{R^2} \right)} \ .
\end{align} 
The spin-orbit contribution can be seen from Eq.~\eqref{Kerr-hamiltonian} to be
\begin{align}
    H_{\rm SO} = \frac{2 M P_\varphi\  a}{R^3 + a^2(R + 2M)}.
\end{align}

For scattering orbits, as the radial action controls the dynamics, it will often be convenient to express the mass-shell condition in terms of the radial momentum and the energy,
\begin{align}
    P_{R}^2 &= \frac{1}{(1 + \Bnp)} \left[ \frac{1}{A^{\rm Kerr}}(H_{\rm Kerr} - H_{\rm SO})^2 \right. \nonumber \\
    &\left. \qquad \qquad \quad - \left( \mu^2 + \frac{P_\varphi^2}{R^2} + \Bnpa \frac{P_\varphi^2 a^2}{R^2} \right) \right].
\end{align}
The scattering angle of a test particle in the Kerr metric can then be derived using
\begin{align}
    \label{eq:scattering_angle_pr}
    \theta + \pi = -2\int_{R_{\min}}^{\infty} dR \, \frac{\partial P_R}{\partial P_\varphi}.
\end{align}
A detailed study of classical scattering in the Kerr spacetime may be found in \cite{Damgaard:2022jem}, where the scattering angle was computed perturbatively to $\mathcal{O}(G^6)$.

An alternative approach to the Kerr Hamiltonian was introduced in~\cite{Damour:2014sva}. 
In this work, a gauge-invariant centrifugal radius $R_c$ was identified, which allows one to factorize the Kerr $A$-potential into the product of a Schwarzschild-like term and a multiplicative correction that encodes the spin-dependent multipolar structure. 
Focusing on just the orbital contribution to the Kerr Hamiltonian, \cite{Damour:2014sva} demonstrated that this can be reexpressed as
\begin{align}
    H^{\rm Kerr}_{\rm orb} = \sqrt{A_{\rm eq} (R_c) \left( \mu^2 + \frac{P_{\varphi}^2}{R_c^2} + \frac{P_R^2}{B_{\rm eq}(R)} \right)},
\end{align}
by introducing the centrifugal radius 
\begin{align}
    R_c^2 &= R^2 + a^2 \left( 1 + \frac{2M}{R} \right).
\end{align}
To establish the connection with our previous notation, and demonstrate the equivalence of these approaches, we provide the following correspondences
\begin{align}
\Lambda(r) &\xrightarrow{\text{eq}} \mathcal{R}^4(r) = R^4 + R^2 a^2 + 2MRa^2\,, \\
A^{\text{Kerr}}(R) &\xrightarrow{\text{eq}} A_{\text{eq}}(R_c) = \frac{\Delta(R) R^2}{\Lambda(R)} = \frac{\Delta(R)}{R_c^2}\,, \\
1 + B(R) &\xrightarrow{\text{eq}} B_{\text{eq}}^{-1}(R) = \frac{\Delta(R)}{R^2}\,, \\
\frac{1}{R^2} + \frac{a^2}{R^2} B_a(R) &\xrightarrow{\text{eq}} \frac{1}{R_c^2}\,.
\end{align}
By inspection, we see that the $A$-potential can be written in a factorized form~\cite{Damour:2014sva}
\begin{align}
    A_{\rm eq}(R_c) &= A_{\rm Schw} (R_c) \hat{A} (R_c), \\ 
    &= \left( 1 - \frac{2M}{R_c} \right)\frac{1 + \frac{2M}{R_c}}{1 + \frac{2M}{R}},
\end{align}
where $A_{\rm Schw}$ is in the usual Schwarzschild-form and $\hat{A} (R_c)$ encodes the multipolar structure of Kerr~\cite{Damour:2014sva}. 
In particular, it was shown in~\cite{Damour:2014sva} that to a low-order PN expansion, the $A$-potential can be expressed as
\begin{align}
A_{\rm PN} (R_c) \approx 1 - \frac{2 M}{R_c} - \frac{M a^2}{R^3_c} - \frac{2 M^2 a^2}{R^4_c} - \frac{3 M a^4}{4 R_c^5} + \mathcal{O}(R_c^{-7}),
\end{align}
where $-2M/R_c$ is the usual monopole term, $-M a^2 / R^3_c$ the quadrupole gravitational potential term seen in the equatorial plane, and so on~\cite{Geroch:1970mom,Hansen:1974mom,Thorne:1980mom,Beig:1981mom,Damour:2014sva}. 
Another subtle point is that the location of the outer-horizon $\Delta (R^+_H) = 0$ in usual Boyer-Lindquist coordinates is highly dependent on the spin, whereas the gauge-invariant equivalent for the centrifugal radius is just $R_c = 2M$~\cite{Damour:2014sva}. 
The reformulation in terms of a centrifugal radius therefore enables us to systematically incorporate spin effects into the EOB Hamiltonian by exploiting the natural factorization of the Kerr potential into a Schwarzschild-like term and multipolar corrections.
Spin-orbit interactions can be included in the usual way through $H_{\rm SO}$.  
This structure will provide a natural starting point for incorporating spin-dependent PM information, as we will see later on. 

Finally, to make contact with previous PN-based work on gauge choices~\cite{Balmelli:2015zsa,Khalil:2023kep}, it is useful to write the Kerr-metric in terms of the following 3-vectors,
\begin{align}
    P_R = (\mathbf{N} \cdot \mathbf{P}), \hspace{2em} P_{\varphi} = (\mathbf{R}  \times \mathbf{P})_{z} = \bfL,
\end{align}
where,
\begin{align}
    \mathbf{P}^2 = P_R^2 + \frac{P_{\theta}^2}{R^2} + \frac{P_{\varphi}^2}{R^2\sin^2{\theta}}.
\end{align}
In this notation, the Kerr metric may be written as,
\begin{align}
   & H_{\rm Kerr} =  \frac{2MR}{\Lambda} \bfL \cdot \bfa \\
   & \notag \qquad + \Big[ \AKerr \Big( \mu^2 + \Bnp (\mathbf{N} \cdot \mathbf{P}) + \frac{R^2}{\Sigma}\mathbf{P}^2 \\
   & \notag \qquad + \Bnpa \left((\mathbf{N} \times \mathbf{P}) \cdot \bfa \right)^2 \Big) \Big]^{\frac{1}{2}}.
\end{align}
The Hamiltonian of a nonspinning test mass in Kerr has many desirable features. For example, the motion is fully integrable, owing to the existence of a hidden symmetry leading to an extra constant of motion, the Carter constant~\cite{Carter:1968con}. 
This is due to the fact that the spacetime admits two killing vectors, e.g. $\partial_t$ and $\partial_\phi$ being two independent choices, but it also admits a Killing-Yano tensor, which directly leads to the Carter constant. 
In addition, the test-mass Hamiltonian naturally encodes all leading even-in-spin PN orders for binary BHs~\cite{Vines:2016qwa} and yields simpler equations of motion relative to Hamiltonians based on the test-spin limit, making it an ideal foundation for constructing EOB Hamiltonians.

\subsection{The post-Minkowskian expansion}
\label{subsection:post-minkowskian}
The basic idea of the PM expansion is to expand the metric in the weak-field regime where $GM / (rc^2) \ll 1$,
\begin{align}
    g_{\mu\nu} &= \eta_{\mu\nu} + h_{\mu\nu}, \qquad  ||h_{\mu\nu}|| \ll 1, \\
    h_{\mu\nu} &= \sum_{n \geq 1} \left(\frac{G}{c^4}\right)^n h^{(n)}_{\mu\nu},
\end{align}
where $\eta_{\mu\nu}$ is the Minkowski metric and $h^{(n)}_{\mu\nu}$ are the $n$th order perturbations with respect to $G$.
The techniques employed in the post-Minkowskian expansion have been readily generalized to include additional physical effects, such as spins~\cite{Bern:2020buy} and tidal effects~\cite{Jakobsen:2023pvx}. In later sections we focus on spin effects, where the current state of the art includes all orders in spin up to 2PM~\cite{Vines:2017hyw,Aoude:2022thd, Aoude:2023vdk}. 
At 3PM, results have been obtained up to $\mathcal{O}(S^4)$ for spin-orbit couplings~\cite{Akpinar:2025bkt} and up to $\mathcal{O}(S^2)$ for the spin-spin couplings~\cite{Jakobsen:2022fcj}. 

A key gauge-invariant observable in the PM expansion is the scattering angle, which takes the form
\begin{align} \label{pm-scattering-angle}
    \theta^{(n \rm PM)} = \pi + \sum_{i=1}^{n} 2 \frac{\theta^{(i)}_{\rm PM}}{p_{\phi}^i},
\end{align}
where $p_{\phi} = P_{\varphi} / (GM\mu)$ plays the role of our PM expansion parameter by tracking factors of $G$ and the coefficients are functions of the energy $\gamma$.
The first two coefficients in the PM-expanded scattering angle are given by~\cite{Westpfahl:1979gu,Westpfahl:1979gu,Damour:2016gwp,Damour:2017zjx}
\begin{align}
    \theta^{(1)}_{\rm PM} &= \frac{2\gamma^2 - 1}{\sqrt{\gamma^2 - 1}}, \\
    \theta^{(2)}_{\rm PM} &= \frac{3\pi}{8}\frac{(5\gamma^2 - 1)}{\sqrt{1 + 2\nu(\gamma - 1)}}.
\end{align}
Radiation reaction enters at 3PM, and the scattering angle becomes comprised of conservative and dissipative contributions~\cite{Amati:1990xe,Damour:2017zjx,Bern:2019nnu,Bern:2019crd,Damour:2019lcq,DiVecchia:2019kta,Kalin:2020fhe}. 
At 4PM, additional complexity arises including nonlocal-in-time tail effects encoded in the special functions appearing in the scattering amplitude~\cite{Bern:2021yeh,Bini:2021gat,Dlapa:2021npj,Manohar:2022dea,Bini:2022enm}. 
The radiation-reacted scattering angle may be deduced from the conservative sector via a linear-response formula~\cite{Bini:2012ji}, depending on the PM results for the radiated energy $E_{\rm rad}$ and angular momentum $J_{\rm rad}$,
\begin{align}
    \theta_{\rm rr} = \frac{1}{2}\frac{\partial \theta^{\rm cons}}{\partial E}\bigg|_{\rm rad} E_{\rm rad} + \frac{1}{2}\frac{\partial \theta^{\rm cons}}{\partial J}\bigg|_{\rm rad} J_{\rm rad}.
\end{align}
Note that this formula has been extended beyond linear responses and for binaries with misaligned spins in \cite{Jakobsen:2022fcj}.

\subsection{Effective one body approach}
\label{subsection:eob}
The EOB formalism~\cite{Buonanno:1998gg, Buonanno:2000ef} has proven to be one of the most powerful tools in understanding two-body dynamics in general relativity. 
The basic idea is to map the full relativistic two-body problem onto the motion of a test mass or test-spin particle in an effective metric, systematically incorporating information from both weak-field perturbation theory and strong-field regimes.  
This approach defines a natural resummation of perturbative information by blending weak-field perturbative information with strong-field results from the test-mass limit, extending results beyond their usual domain of validity. 
We focus on relativistic scattering of black holes, and will seek to include post-Minkowskian information into the EOB framework building on recent studies~\cite{Damour:2016gwp,Damour:2017zjx,Damour:2019lcq,Antonelli:2019ytb,Damour:2022ybd,Khalil:2022ylj,Rettegno:2023ghr,Buonanno:2024byg,Buonanno:2024byg,Damour:2025uka}.

We use the following rescaled dimensionless coordinates and momenta,
\begin{align}
    r = \frac{R}{GM}, \quad t = \frac{T}{GM}, \quad \phi = \frac{\varphi}{GM}, \quad p_\mu &= \frac{P_\mu}{\mu}.
\end{align}
where $(R,T)$ are the physical coordinates and $P_\mu$ are the physical canonical momenta. 
The angular momentum $p_{\phi}$ is related to the total angular momentum as
\begin{align}
    p_{\phi} = \frac{P_{\varphi}}{G \mu M}.
\end{align}
For spinning black holes with Kerr parameters $a_1$ and $a_2$, we further define the following useful spin variables
\begin{align}
    a_\pm &= a_{1} \pm a_2, \\ 
    \chi_{\pm} &= \frac{a_{\pm}}{GM},
\end{align} 
where $\chi_\pm$ are dimensionless spin variables.

The dynamics of a test particle in a generic EOB effective metric are governed by the mass shell condition, which takes the following form in our rescaled variables
\begin{align}\label{mass-shell-eob}
    1 + g_{\rm eff}^{\mu\nu}p_{\mu} p_{\nu} + Q(r, \gamma, \nu) = 0,
\end{align}
where $Q = \bar{Q} / \mu^2$ is a nongeodesic term that can be used to incorporate PM deformations. 
One can derive an effective Hamiltonian by solving for $\hat{H}_{\rm eff} = -p_{0}$, which is related to the real two body dynamics via the EOB energy map~\cite{Buonanno:1998gg}
\begin{align} \label{energy-map}
    H_{ \rm EOB } = M\sqrt{1 + 2\nu \left (\frac {H_{ \rm eff }}{\mu}-1 \right ) }.
\end{align}
We note that $H_{\rm EOB} = E_{\rm real} = \sqrt{s}$ for a scattering trajectory, where $s$ is the usual Mandlestam invariant.
From this, we can define the following dimensionless energy variables
\begin{align}
    \Gamma &= \frac{E_{\rm real}}{M} = \sqrt{1 + 2\nu(\gamma - 1)}, \\
    \gamma &= \frac{H_{\rm eff}}{\mu}, 
\end{align}
where $H_{\rm eff}$ is the effective Hamiltonian and $E_{\rm real}$ is the center-of-mass (CoM) energy of the real two-body system.
In the initial CoM frame, the incoming momenta are just $p^{\mu}_1 = (E_1, \boldsymbol{p}_{\infty} / \Gamma)$ and $p^{\mu}_2 = (E_2, - \boldsymbol{p}_{\infty} / \Gamma)$ such that $\gamma = (p_1 \cdot p_2) / (m_1 m_2)$, e.g. as discussed in~\cite{Damour:2022ybd,Buonanno:2024vkx}.
We note that conventions for these quantities can vary significantly across the literature.

The EOB approach has traditionally been used to resum PN information by applying canonical transformations to map PN Hamiltonians into EOB coordinates, in conjunction with the energy map in Eq.~\eqref{energy-map}.
However, PM information derived from scattering calculations has a fundamentally more intricate structure, with energy-dependent coefficients and complex mass-ratio dependence~\cite{Damour:2016gwp}. This makes the transcription of PM information into the EOB framework technically challenging.
A key breakthrough was the identification that one can use the gauge-invariant scattering angle, defined in Eq.~\eqref{pm-scattering-angle}, to determine the coefficients in the EOB model~\cite{Damour:2016gwp, Damour:2017zjx, Damour:2022ybd}.
The fundamental relationship is the matching condition, which requires the EOB-predicted scattering angle reproduce the known PM results at a given PM order,
\begin{align} \label{angle-matching}
    \theta^{\rm (EOB)}_{n \rm PM} = \theta^{(n \rm PM)}.
\end{align}
This allows us to determine any unknown coefficients in the EOB effective potential, thereby determining the complete EOB dynamics.

A foundational problem when applying PM information to bound-orbit models is that the PM expansion coefficients depend on the total energy of the system, which is itself determined by the Hamiltonian.
Recent work has tried to tackle this recursive definition in a number of ways.
In the nonspinning limit, one can just express $\gamma$ in terms of the Schwarzschild Hamiltonian $H_{\rm Schw}$ plus PM corrections~\cite{Damour:2017zjx,Antonelli:2019ytb,Khalil:2022ylj}.
This philosophy was also adopted in~\cite{Buonanno:2024byg}, whereby $\gamma$ was interpreted as an effective energy that is PM-expanded around the Kerr background, $\gamma \sim \gamma_{\rm Kerr} + \sum_n \Delta^{(n)} (\gamma_{\rm Kerr})$, with $\Delta^{(n)}$ corresponding to energy-dependent PM deformation coefficients determined to a given PM order.
An alternative framework to tackle this problem was recently proposed in~\cite{Damour:2025uka} using a novel Lagrange-multiplier based approach for constructing EOB-PM models, aptly called the LEOB model.  
In particular, it was shown that by incorporating the mass-shell condition as a constraint via a Lagrange multiplier, one can introduce an evolution equation for $\gamma = \hat{H}_{\rm eff}$ and explicitly solve for the energy dependence~\cite{Damour:2025uka}. 
In this work, we only focus on the transcription of PM information into Hamiltonians for unbound scattering orbits, thereby avoiding the issue. 
In the subsequent sections we will outline the current approaches to constructing EOB-PM orbits, before exploring the impact of gauge-flexibility on the accuracy of predictions for the scattering angle within the EOB framework.

\subsection{The \texorpdfstring{$w_{\rm eob}$}{w-eob} and SEOB-PM models} \label{subsection:seob+weob}
In this manuscript, we focus on two closed-form EOB-PM models for gravitational scattering: i) the $w_{\rm eob}$ model introduced in \cite{Damour:2022ybd, Rettegno:2023ghr}, and ii) the SEOB-PM model introduced in~\cite{Buonanno:2024byg, Buonanno:2024vkx}. 
These models allow us to directly calculate the scattering angle via Eq.~\eqref{prsq-PM}.
We note that we could also have explored EOB models that integrate the full dynamics together with some prescription for incorporating radiation reaction effects, for example TEOBResumS~\cite{Nagar:2018zoe,Hopper:2022rwo,Albanesi:2024xus,Albanesi:2025txj} or SEOBNRv5~\cite{Khalil:2023kep,Ramos-Buades:2023ehm,Long:2025nmj}. 
We leave such an exploration to future work.

The two EOB-PM models differ in several key ways, with the various choices leading to differing predictions for the scattering angle across the parameter space. Here we outline some of the most relevant differences. 
First, regarding gauge, SEOB-PM employs standard Schwarzschild coordinates whereas $w_{\rm eob}$ uses isotropic coordinates. 
More significantly, each model implements PM deformations through distinct gauge approaches. 
In addition, SEOB-PM adopts the post-Schwarzschild${\ast}$ (PS$\ast$) gauge~\cite{Antonelli:2019ytb,Khalil:2022ylj}, where PM corrections enter directly through deformations of the metric potentials $A$ and $g_{a\pm}$, obeying the mass-shell constraint $g_{\mu\nu}^{\rm eff} P^\mu P^\nu = -\mu^2$. 
In contrast, $w_{\rm eob}$ employs the post-Schwarzschild (PS) gauge~\cite{Damour:2016gwp,Damour:2017zjx}, where the metric potentials remain fixed to the Schwarzschild background and PM information enters exclusively via the non-geodesic term $Q$, leading to a mass-shell condition of the form $g_{\mu\nu}^{\rm eff} P^\mu P^\nu + \bar{Q} = -\mu^2$.
Second, $w_{\rm eob}$ employs PM-expanded EOB radial potentials, $w \sim \sum_n w_n \bar{r}^{-n}$, which directly determines the radial momentum structure and scattering angle~\cite{Damour:2022ybd,Rettegno:2023ghr}. 
SEOB-PM instead uses the impetus formula to define an effective resummation of the $w$-potential~\cite{Buonanno:2024vkx}.
Finally, they differ in terms of their behavior in the test-mass limit, a direct consequence of the gauge choices. 
In particular, $w_{\rm eob}$ does not reduce to the exact geodesic limit but rather a PM-truncated approximation, see Fig.~9 of~\cite{Buonanno:2024vkx}, whereas SEOB-PM explicitly recovers the exact test-mass limit by construction. 
See also~\cite{Damour:2025uka} for a detailed discussion of these issues in the context of the LEOB model.
 
Regarding spin, the SEOB-PM model employs the Kerr metric in Boyer-Lindquist coordinates, Eq.~\eqref{Kerr-metric}, as the foundation for the EOB Hamiltonian. 
Even-in-spin deformations enter through the $A$-potential in the orbital part of the Hamiltonian, while odd-in-spin deformations are incorporated via the gyrogravitomagnetic couplings in the spin-orbit sector.
In contrast, the current spinning-sector work of the $w_{\rm eob}$ model takes a different approach and introduces a spin-expansion of the w-potential coefficients to include the spin information. Due to the fact that the SEOB-PM type models recover the full Kerr Hamiltonian in the test mass limit, we will follow and take that approach to explore the impact of gauge choices for spinning-EOB Hamiltonians. 

In the following sections we will construct a variety of SEOB-PM type models with different gauge choices, in both the nonspinning and spinning sectors. 
By comparing the scattering angles computed in these models with numerical relativity we assess the impact of these gauge choices on the performance of the model alongside the specific resummation of the radial momenta.

\section{The nonspinning sector} 
\label{section:Non-Spinning}
We begin by studying EOB-PM models in the nonspinning limit, describing the motion of a test mass particle in an deformed Schwarzschild spacetime. 
The dynamics are governed by the mass-shell condition
\begin{align}
    1 + g_{\rm eff}^{\mu\nu} p_\mu p_\nu + Q(r,\gamma, \nu) = 0,
\end{align}
together with a deformed (effective) Schwarzschild metric~\cite{Wald:1984rg}
\begin{align}
    g^{\mu\nu}_{\rm eff}\pd_\mu \pd_\nu = -\frac{1}{A(r,\gamma, \nu)}\pd_t^2 + \frac{1}{B(r)}\pd_r^2 + \frac{1}{C(r)}\pd_\phi^2.
\end{align}
By construction, the potentials are consistent with the Kerr metric potentials in the nonspinning limit, i.e. 
\begin{align}
    A^{\rm Kerr}(r) \xrightarrow{a = 0} A(r), \quad
    1 + \Bnp \xrightarrow{a = 0} B(r).
\end{align}
The term $Q(r,\gamma, \nu)$ is a nongeodesic term altering the mass-shell condition, which may be used to incorporate PM deformations. The general form of the radial momentum is found by solving this mass-shell condition for $p_r$, 
\begin{align}\label{radial-momentum}
    p_r^2 = \frac{\gamma^2 B(r)}{A(r, \gamma, \nu)} - \frac{\pphi^2 B(r)}{C(r)} - B(r) - B(r)Q(r,\gamma, \nu).
\end{align}
We could also solve for the effective Hamiltonian, $H_{\rm Schw} = \gamma$,
\begin{align}
    H_{\rm Schw} = \sqrt{A(r) \left( 1 + \frac{\pphi^2}{r^2} + \frac{p_r^2}{B(r)} \right)}.
\end{align}

In the nonspinning sector, there are several gauge choices that must be addressed. 
Due to the spherically symmetric nature of the problem, one key choice is the definition of radial coordinate. 
In this work, we consider the standard Schwarzschild radial coordinate $r$ as well as isotropic coordinates $\bar{r}$, where the latter is defined by
\begin{align}
    \bar{C}(\bar{r}) = \bar{r}^2\bar{B}(\bar{r}),
\end{align}
where an overbar denotes variables derived in isotropic coordinates. 
In both sets of coordinates, the metric potentials are given by
\begin{align}
    A(r) &= 1 - \frac{2}{r}, & \bar{A}(\bar{r}) &= \left( \frac{1 + \frac{1}{2\bar{r}}}{1 - \frac{1}{2\bar{r}}} \right)^2, \\
    B(r) &= \frac{1}{A(r)}, & \bar{B}(\bar{r}) &= \left( 1 + \frac{1}{2\bar{r}} \right)^4, \\
    C(r) &= r^2, & \bar{C}(\bar{r}) &= r^2 \bar{B}(\bar{r}).
\end{align}
The second main choice of gauge concerns how to incorporate perturbative information into the EOB metric potentials~\cite{Buonanno:1998gg,Buonanno:2000ef,Damour:2000we,Damour:2001tu}. 
In the nonspinning sector, the two main gauge choices explored in the literature are post-Schwarzschild (PS) and post-Schwarzschild* (PS*)~\cite{Akcay:2012ea, Damour:2017zjx, Antonelli:2019ytb, Khalil:2022ylj}.
Building on complementary studies in the literature~\cite{Damour:2019lcq,Antonelli:2019ytb,Khalil:2022ylj,Damour:2025uka}, we explore the impact of these gauge choices on the scattering angle in the following sections.

\subsection{General gauge choices}
\label{subsection:general-gauge}
To start we will consider the general form of the mass-shell condition (\ref{mass-shell-eob}), where $A(r)$, $B(r)$ and $Q(r)$ are now taken to be PM expansions in terms of general coefficients
\begin{subequations}
\begin{align}
    A(r) &= 1 + \sum_{i\geq1} \frac{a_{i}(\gamma, \nu)}{r^i}, \\
    B(r) &= 1 + \sum_{i\geq 1} \frac{b_i(\gamma, \nu)}{r^i}, \\
    Q(r) &= \sum_{i\geq 2}\frac{q_{i}}{r^i}.
\end{align}    
\end{subequations}
We take the nongeodesic term $Q$ to start from 2PM, as it was shown that to $\mathcal{O}(G)$ the scattering angle is completely determined by the test-mass limit and hence $Q \sim G^2 + G^3 + \mathcal{O}(G^4)$~\cite{Damour:2017zjx}. 
Likewise, we fix $C$ to the Schwarzschild background, such that $C(r)=r^2$. Note that while we could have chosen to deform the $C$-potential, we keep it fixed to maintain consistency with the conventions across different EOB models.

To compute the PM deformation coefficients, we match the EOB scattering angle to the gauge-invariant PM scattering angle. 
In the nonspinning limit, complete $4$PM information is available together with $5$PM up to $1$SF. 
As the $5$PM sector is as-of-yet incomplete, we include information up to $4$PM for the main discussion before introducing the incomplete $5$PM information.

As previously shown in \cite{Buonanno:2024byg, Damour:2022ybd}, the scattering angle is computed by PM expanding the usual formula from Hamilton-Jacobi theory
\begin{align}
    \theta + \pi = -2\int_{r_{\rm min}}^{\infty} dr\  \frac{\pd p_r}{\pd \pphi},
\end{align}
where $r_{\rm min}$ is the largest positive root of $p_r = 0$. 
As the lower limit of integration depends on the PM order, explicitly PM expanding the integral would be arduous. 
Therefore it is useful to employ the prescription from \cite{Damour:1988mr} in which the lower limit of the integral is taken to be its leading order value, then the integrand is PM expanded and the \textit{partie finie} of the resulting (divergent) integral is taken: 
\begin{align}\label{finite-part-integral}
    \theta^{(n \rm PM)} + \pi = -2\  {\rm Pf} \int_{r_{\rm min}}^{\infty} dr\ E_{n \rm PM}\left[ \frac{\pd p_r}{\pd \pphi} \right],
\end{align}
where $E_{n \rm PM}[\cdot]$ tells us to PM expand the integrand to order $n$. This allows for the efficient calculation of the integrals to arbitrarily high order.
After a suitable change of variables and subtracting off the leading order value, the integrals are all of the form,
\begin{align} \label{eq:partie-finie-beta}
    \theta^{(n \rm PM)} \sim 2\ {\rm{Pf}} \int_{0}^{1} dz (1-z)^{-\frac{1}{2}-n} z^m. 
\end{align}
This is explicitly in the form of a Euler-Beta function, where it is simple to take the finite part, as the analytic structure is well understood.

Substituting our PM-expanded ansatz for the different metric potentials into the radial momentum (\ref{radial-momentum}), we compute the finite part integral of Eq.~\eqref{finite-part-integral}. We fix the $C$-potential, for ease, to its Schwarzschild value.  
This allows the computation of general relations between the undetermined coefficients. 
The first relation is given by
\begin{align}\label{1pm-general-gauge}
    \gamma^2(4 + a_1(\gamma,\nu) - b_1(\gamma, \nu)) - 2 + b_1(\gamma, \nu) = 0.
\end{align}
These coefficients may be fixed by the requirement that the $1 \rm PM$ EOB potentials agree with the PM expansion of the test mass limit. 
For example, by matching the potentials in Schwarzschild coordinates, $a_1$ and $b_1$ are found to be
\begin{align}
    a_1(\gamma, \nu) = -2, \\
    b_1(\gamma, \nu) = 2,
\end{align}
which automatically satisfies Eq.~\eqref{1pm-general-gauge} without need for extra deformation terms. 
The second relationship then simplifies to
\begin{align}
    &\frac{3 - 15\gamma^2}{\Gamma} + 1 - b_2(\gamma, \nu) \\
    &\hspace{2em} + \gamma^2(11 - 2a_2(\gamma, \nu) + b_2(\gamma, \nu)) + 2q_2(\gamma, \nu) = 0. \notag
\end{align}
As we can see, there is a large amount of flexibility, with the test-limit only determining each coefficient up to $\nu$-dependent quantities.
Notice in particular that we can use the test-mass limit to fix all but one of the deformation terms, meaning that, in general, we will only a single deformation coefficient in the nonspinning EOB model.
The choices which are made will greatly affect the accuracy of the overall model, so care needs to be taken when determining the gauge coefficients. 
In the next section we will look at a particular class of EOB gauge, which forms the basis of the current PM-informed EOB models. 

\subsection{Post-Schwarzschild gauges}
\label{subsection:post-schwarzschild}
In the PS gauge the effective metric is taken to be the Schwarzschild metric. Then, all deformations are included in a nongeodesic term $Q$. The mass-shell condition then becomes,
\begin{align}
    1 + g_{\rm schw}^{\mu\nu}p_{\mu}p_{\nu} + Q(r,\gamma, \nu) = 0.
\end{align}
On the other hand, the PS* gauge sets $Q=0$ and incorporates PM information via deformations of the metric potential, 
\begin{align}
    & 1 + g^{\mu\nu}_{\rm eff} p_{\mu}p_{\nu} = 0, \\
    & g^{\mu\nu}_{\rm eff}\pd_{\mu}\pd_{\nu} = \frac{1}{A_{\rm eff}(r,\gamma,\nu)}\pd_t^2 + \frac{1}{B(r)}\pd_r^2 + \frac{1}{r^2}\partial_\phi^2.
\end{align}
For the nonspinning case, spherical symmetry allows us to consider purely equatorial orbits without loss of generality.

Both the $Q$-term and the deformed $A$-potential admit PM expansions, with deformations beginning at $2$PM since the test-mass limit fully determines the dynamics through 1PM~\cite{Damour:2016gwp}.
Specifically, we have 
\begin{align}
    Q(r, \gamma, \nu) &= \sum_{i \geq 2} \frac{q_i(\gamma,\nu)}{r^i}, \\
    A_{\rm eff}(r,\gamma,\nu) &= 1 - \frac{2}{r} + \Delta A,
\end{align}
where the deformation takes the form
\begin{align}
    \Delta A = \sum_{i\geq2}\frac{\alpha_{i}(\gamma,\nu)}{r^i}.
\end{align} 
Following \cite{Buonanno:2024vkx, Damour:2022ybd}, we relate the effective $w$-potential to the radial momentum through the impetus formula
\begin{align}\label{impetus-formula}
    p_r^2 = (\gamma^2 - 1) - \frac{p_\phi^2}{r^2} + w(r,p_\phi, \gamma, \nu), 
\end{align}
which allows us to express $w$ explicitly as~\cite{Damour:2017zjx}
\begin{align}
    \label{w-potential-definition}
    w(r, p_\phi, \gamma, \nu) &= \left( 1 - \frac{B(r)}{C(r)} \right)\frac{p_\phi^2}{r^2} +  \left( \frac{B(r)}{A(r)} - 1 \right)\gamma^2 \\ 
    \nonumber &\qquad \qquad - (B(r) - 1) - B(r)Q(r).
\end{align}
By construction, this admits a PM expansion of the form
\begin{align}\label{w-pm-expansion}
    w(r,p_\phi,\gamma,\nu) = \sum_{i\geq 1}\frac{w_i(p_\phi,\gamma,\nu)}{r^i}.
\end{align}
An advantage of isotropic coordinates is that the $w$-potential simplifies considerably and typically becomes independent of $p_\phi$. However, rather than PM-expanding the $w$-potential, one could also use Eq.~\eqref{impetus-formula} to directly define a resummed potential. 
We will examine both PM-expanded and resummed forms of the $w$-potential, as they provide complementary insights depending on the gauge choice.
As such, the two forms of the radial momentum that we will consider are
\begin{align}\label{prsq-PM}
    p^2_{r, \rm PM} &= (\gamma^2 - 1) - \frac{\pphi^2}{r^2} + \sum_{i \geq 1} \frac{w_i(\pphi, \gamma, \nu)}{r^i}, \\
    \label{prsq-resum}
    p^2_{r, \rm resum} &= (\gamma^2 - 1) - \frac{\pphi^2}{r^2} + w(\pphi, \gamma, \nu),
\end{align}
where $w(\pphi, \gamma, \nu)$ is defined by Eq. \eqref{impetus-formula}.

\subsubsection{The post-Schwarzschild gauge}

We begin by constructing the post-Schwarzschild effective metric to the $4$PM order. In the following we consider two coordinate gauges, in particular isotropic coordinates (as used in \cite{Damour:2022ybd}) and Schwarzschild coordinates (as used in \cite{Buonanno:2024vkx}). Explicitly, the solution of the mass-shell condition for $p_r^2$ is given in each coordinate system by: 
\begin{subequations} \label{isotropic-prsq-ps}
    \begin{align}
    p_{r, (\rm iso)}^2 &= \frac{\bar{B}(\bar{r})}{\bar{A}(\bar{r})}\gamma^2 - \frac{\pphi^2}{\bar{r}^2} - \bar{B}(\bar{r}) - \bar{B}(\bar{r})\bar{Q}(\bar{r}), \\
    p_{r,(\rm schw)}^2 &= \frac{B(r)}{A(r)}\gamma^2 - \frac{B(r)}{C(r)}\pphi^2 - B(r) - B(r)Q(r)
\end{align}
\end{subequations}
There are two possibilities for computing the coefficients of the $w$-potentials defined in Eq.~\eqref{w-pm-expansion}. 
One can either substitute the PM expanded impetus formula, in terms of undetermined $w$-coefficients and compute them directly from the scattering angle, or substitute Eq.~\eqref{isotropic-prsq-ps} into Eq.~\eqref{impetus-formula} and PM expand to find a direct relationship between the coefficients $\bar{q}_i(\gamma, \nu)$ and $w_i(\gamma, \nu)$. 
Taking the latter route allows us greater flexibility in defining the model and in particular gives us both the PM-expanded $w_{\rm eob}$ type models and the resummed SEOB-PM type models. 

The relationships for the $w_i(\gamma, \nu)$ coefficients in terms of $q_i(\gamma, \nu)$ are given by the following formulae, found by PM expanding Eqs.~\eqref{isotropic-prsq-ps}. The isotropic coordinate expressions are given by:
\begin{subequations} \label{w-potential-coefs-ps-isotropic}
\begin{align}
    \bar{w}_1(\gamma, \nu) &= 2 \left(2 \gamma ^2-1\right), \\
    \bar{w}_2(\gamma, \nu) &= \frac{1}{2}(-3 + 15\gamma^2 - 2\bar{q}_2(\gamma, \nu)), \\
    \bar{w}_3(\gamma, \nu) &= -\frac{1}{2} + 9\gamma^2 - 2\bar{q}_2(\gamma, \nu) - \bar{q}_3(\gamma, \nu), \\
    \bar{w}_4(\gamma, \nu) &= \frac{1}{16}(-1 + 129\gamma^2 - 24\bar{q}_2(\gamma, \nu) \\
    &\hspace{2em} - 32\bar{q}_3(\gamma, \nu) - 16\bar{q}_4(\gamma, \nu)), \notag
\end{align}
\end{subequations}
and for Schwarzschild coordinates,
\begin{subequations} \label{w-potential-coefs-ps-schwarzschild}
    \begin{align}
        w_1(\gamma, \nu) &= 2(2\gamma^2 - 1), \\
        w_2(\gamma, \nu) &= -4 + 12\gamma^2 - q_2(\gamma, \nu), \\
        w_3(\gamma, \pphi, \nu) &= -8 - 2\pphi^2 + 32\gamma^2 - 2q_2(\gamma, \nu) - q_3(\gamma, \nu),  \\
        w_4(\gamma, \pphi, \nu) &= -16 - 4\pphi^2 + 80\gamma^2 - 4q_2(\gamma, \nu), \\
        &\hspace{2em} - 2q_3(\gamma, \nu) - q_4(\gamma, \nu). \notag 
    \end{align}
\end{subequations} 
In isotropic coordinates the $w$-potentials depend only on the energy and symmetric mass ratio, whereas in Schwarzschild coordinates they have a more complicated dependence on the angular momentum $\pphi$.
Note that the $w_{\rm eob}$ model of Ref.~\cite{Damour:2022ybd} employs isotropic coordinates, and our expressions using the deformation coefficients in Appendix~\ref{appendix-deformation-coefs-no-spin} agree with their construction, which calculates deformations through an alternative method.

The deformation coefficients $q_i$ are computed using Eq.~\eqref{finite-part-integral} with the radial momentum from Eq.~\eqref{isotropic-prsq-ps}. 
Following the procedure in~\cite{Damour:1988mr,Damour:2016gwp}, we PM-expand the resulting scattering angle integral and apply the Hadamard \textit{partie finie}, see Eq.~\eqref{eq:partie-finie-beta}, to derive relations between the $q_i$ coefficients and the scattering angle. 
The coefficients $q_i(\gamma, \nu)$ are then determined by matching the PM-expanded EOB scattering angle to known post-Minkowskian results using Eq.~\eqref{angle-matching}.
The computed deformation coefficients are listed in Appendix \ref{appendix-deformation-coefs-no-spin}. 

With the $q_i$ coefficients at hand, the effective Hamiltonian $H_{\rm eff} = \mu \gamma = -\mu p_0$ can be obtained from the mass-shell condition (Eq.~\eqref{mass-shell-eob}) for use in EOB models.

\subsubsection{The post-Schwarzschild\texorpdfstring{$\ \ast$}{*} gauge}
In the post-Schwarzschild star (PS*) gauge, the test-body motion must be geodesic in a deformed Schwarzschild metric. 
We therefore incorporate PM information directly into the metric potentials by fixing $B(r)$ and $C(r)$ and deforming the $A$-potential with $\nu$-dependent coefficients.
The mass shell condition is given in both coordinate systems by
\begin{subequations}
    \begin{align} \label{radial-momentum-PS*}
        p_{r,(\rm iso)}^2 &= \frac{\bar{B}(\bar{r})}{\bar{A}(\bar{r},\gamma,\nu)}\gamma^2 - \frac{\pphi^2}{\bar{r}^2} - \bar{B}(\bar{r}), \\
        p_{r,(\rm schw)}^2 &= \frac{B(r)}{A(r, \gamma, \nu)}\gamma^2 - \frac{B(r)}{C(r)}\pphi^2 - B(r). 
    \end{align}
\end{subequations}
Here the metric potential $A(r, \gamma, \nu)$ includes deformation coefficients $\alpha^{(n)}$ (this notation aligns with \cite{Buonanno:2024vkx}, without the spin labels). In Schwarzschild coordinates this takes the following form,
\begin{align}
    A(r, \gamma, \nu) = 1 - \frac{2}{r} + \Delta A \\
    \Delta A = \sum_{n \geq 2} \frac{\alpha^{(n)}(\gamma, \nu)}{r^n}.
\end{align}
For isotropic coordinates, we follow the same principle but using the isotropic potential instead of Schwarzschild. 
As before, we denote the isotropic deformation coefficients as $\bar{\alpha}_i(\gamma, \nu)$.

We may again define PM expanded w-potentials from the impetus formula, Eq.~\eqref{impetus-formula}. The isotropic gauge w-potential coefficients take the form,
\begin{subequations}
\begin{align}
    \bar{w}_1(\gamma) &= 2(2\gamma^2 - 1), \\
    \bar{w}_2(\gamma, \nu) &= \frac{1}{2}(\gamma^2(15 - 2\bar{\alpha}_2(\gamma, \nu)) - 3), \\
    \bar{w}_3(\gamma, \nu) &= -\frac{1}{2} -\gamma^2(\bar{\alpha}_3(\gamma, \nu) + 6\bar{\alpha_2}(\gamma, \nu) - 9), \\
    \bar{w}_4(\gamma, \nu) &= -\frac{1}{16} + \gamma^2 \Big( \frac{129}{16} - \frac{35}{2}\bar{\alpha}_2(\gamma, \nu), \\
    &\hspace{2em} + \bar{\alpha}_2(\gamma, \nu)^2 - 6\bar{\alpha}_3(\gamma, \nu) - \bar{\alpha}_4(\gamma, \nu) \Big). \notag 
\end{align}
\end{subequations}
The Schwarzschild gauge potential coefficients are given by,
\begin{subequations}
\begin{align}
    w_1(\gamma) &= 2(2\gamma^2 - 1), \\
    w_2(\gamma, \nu) &= -4 -\gamma^2 ( -12 + \alpha_2(\gamma, \nu)), \\
    w_3(\gamma, \nu, \pphi) &= -8 - 2\pphi^2 + \\
    &\hspace{2em} \gamma^2\Big( 32 - 6\alpha_2(\gamma, \nu) - \alpha_3(\gamma, \nu) \Big), \notag \\
    w_4(\gamma, \nu, \pphi) &= -16 - 4\pphi^2 + \gamma^2 \Big( 80 -24\alpha_2(\gamma, \nu), \\
    &\hspace{2em} + \alpha_2(\gamma, \nu)^2   - 6\alpha_3(\gamma, \nu) - \alpha_4(\gamma, \nu) \Big). \notag
\end{align}
\end{subequations}
The calculation of the deformation coefficients uses exactly the same methodology as for post-Schwarzschild, and they may be found in Appendix~\ref{appendix-deformation-coefs-no-spin}.

\subsubsection{Structural differences between the models}

With the deformation coefficients and $w$-potential mappings in hand, we now investigate the differences between the various models. At 2PM, the coefficients may be deduced from the test-mass limit, with $\nu$-dependence starting due to the EOB-energy map, see Eq. (2.32) of \cite{Damour:2019lcq}. In this case the coefficients take the simple form,
\begin{align}
    \bar{q}_2(\gamma,\nu) &= \frac{3}{2}(5\gamma^2 - 1) \left( \frac{1}{\Gamma} - 1 \right).
\end{align}
The only difference between PS and PS* here is an extra factor of $\gamma^2$, due to the different ways that the information is incorporated into the effective metric. 
At 3PM the structure of the coefficients begins to vary across the gauge choices, and we include explicit expressions in Appendix ~\ref{appendix-deformation-coefs-no-spin}. 
At $4$ and $5$PM the expressions increase in complexity, including elliptic functions and polylogarithms, inherited from the scattering angles, and we include explicit expressions in the ancillary file.  
In the following comparisons between the different gauge choices, we will work at 4PM, unless otherwise stated.

It is important to note that, while the coefficients are different between PS and PS*, the coefficients of the $w$-potentials also differ. We can explicitly show that for the PM expanded radial momentum (as in Eq.~\eqref{prsq-PM}), the choice of PS or PS* yields an identical expression. 
This can be seen by comparing the effective potentials and scattering angles, which we present below below in Fig.~\ref{fig:potentials-all-gauges-pm-expansion} and Fig.~\ref{fig:Expanded-vs-resummed} respectively. 

Differences between the two EOB gauges occur when we consider the resummed radial momentum, as in Eq.~\eqref{prsq-resum}, employed in for example, \cite{Buonanno:2024vkx}. 
The different forms of the resummed $w$-potential drastically change the form of the radial momentum and therefore the accuracy of the model.
It is instructive to study the structure of the effective potentials in the different gauges, which we define as the following:
\begin{align}
    V_{\rm eff} = \frac{\pphi^2}{r^2} - w(r, \pphi, \gamma, \nu).
\end{align}

\begin{figure*}[!ht]
    \centering 
    \includegraphics[width=\linewidth]{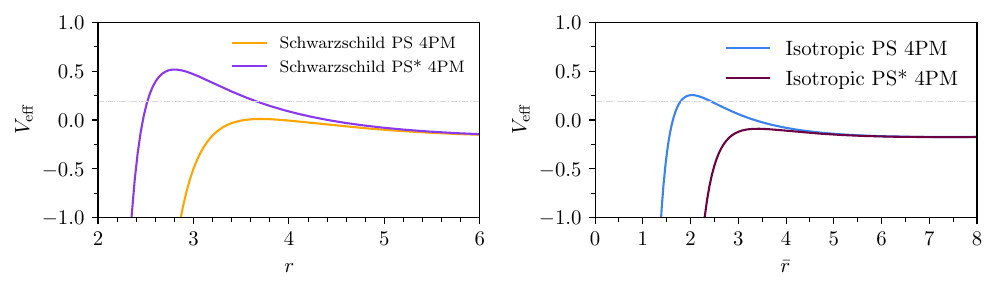}
    \caption{A comparison of the effective potentials at 4PM for an equal mass configuration at fixed energy ($\Gamma_1 \approx 1.02$) and angular momentum ($\pphi \approx 4.4$). The left panel compares the two EOB gauges for Schwarzschild coordinates and the right for isotropic coordinates.
    \label{seob-type-potentials-4PM}
    }
\end{figure*}

Figure \ref{seob-type-potentials-4PM} shows the 4PM accurate resummed effective potentials across the various choices of gauge. For reference, the Schwarzschild PS* curve corresponds to the SEOB-PM model. The horizontal line corresponds to $p_\infty^2 = \gamma^2-1$, which is the dimensionless initial energy of the effective test mass. 
If the curve remains below $p_{\infty}^2$ for all radii, then the model always predicts a plunge at this value of angular momentum, whereas if the curve goes above it predicts a scatter. 

\begin{figure*}
    \centering
    \includegraphics[width=\linewidth]{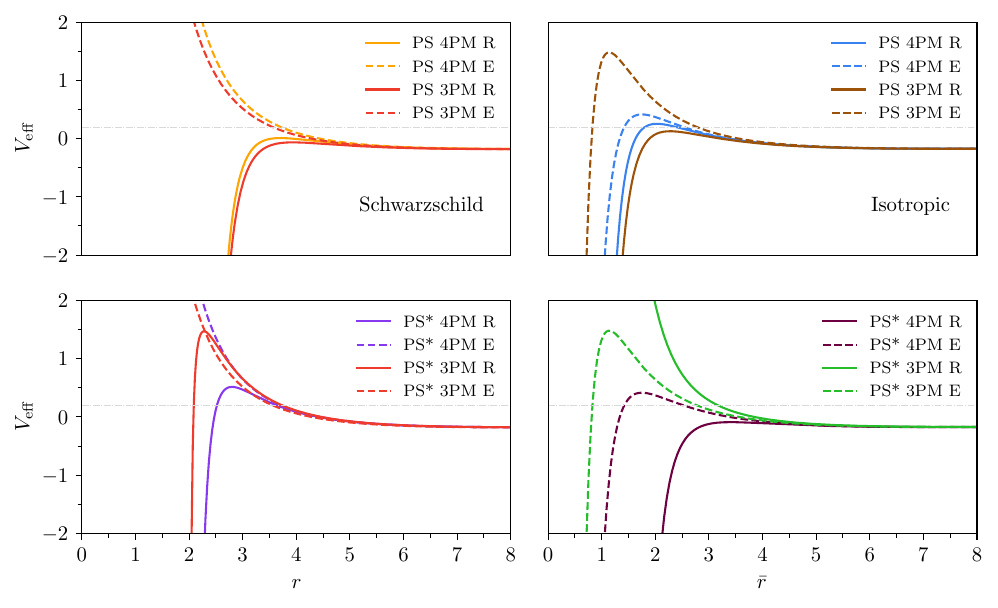}
    \caption{A comparison of effective potentials across the different PM orders considered in all combinations of gauge choices. The left panels are the potentials with a Schwarzschild radial coordinate, and the right have an isotropic radial coordinate. Note that the PM expanded potentials ($w_{\rm eob}$ type) in PS and PS* are equivalent. Note also the similarity between the PS* resummed (SEOB-PM model) and the isotropic PM expanded potentials ($w_{\rm eob}$ model). All figures have energy $\Gamma_1$ and dimensionless angular momentum $\pphi\approx4.4$. The dashed line is the dimensionless quantity $p_{\infty}^2 = \gamma^2 - 1 $, if the potential never rises above the line then that model will always plunge for this value of $l$. Numerical relativity simulations at this energy and angular momentum \cite{Swain:2024ngs} show that the system scatters.
    \label{fig:potentials-all-gauges-pm-expansion}
    }
\end{figure*}

We can see from the structure of the potentials that the gauge choice has a large impact on whether the model predicts a plunge. 
Figure \ref{fig:potentials-all-gauges-pm-expansion} shows the impact of PM expansion on the effective potentials across the gauges. It is notable that the comparison between resummed and PM expanded strongly depends on the coordinate gauge as well as the EOB gauge. 
This comparison also shows that for PM expanded potentials, PS and PS* are equivalent. An additional comparison showing the impact of PM order may be found in Fig \ref{fig:figure-8} This is also be evident from the scattering angle comparisons in Figs.~\ref{fig:PSvsStar-isotropic} and \ref{fig:PSvsStar-schwarzschild}. 
Notice that for the isotropic case the shapes of the PM expanded potentials are similar to the resummed Schwazschild potentials, while the resummed isotropic potentials have a very different structure, with both 4PM isotropic effective potentials predicting a plunge at this angular momentum. 
We can see from the potential comparisons that the resummed isotropic potential with a PS* gauge has unexpected results, with the $3$PM potential diverging, predicting a scatter but the $4$PM potential has the opposite behavior. 
A similar divergent behavior is identified in the PM expanded Schwarzschild coordinate potentials. 
The expected behavior of the potentials, stemming from intuition of the test-mass limit, is only seen in the Schwarzschild PS* and PM expanded isotropic potentials. 

Given the deformation coefficients it is simple to construct the Hamiltonian in each gauge. It is important to note that starting at the 4PM order there are nonlocal terms which enter the deformation coefficients, invalidating the current scattering to bound analytic continuation. 
There are prescriptions available to remedy this, such as the so called Tutti-Frutti approach~\cite{Bini:2019nra, Bini:2020hmy}, which incorporates information from a different perturbative approaches to compute the local and nonlocal parts of the radial action. 
This formalism has recently been applied in the LEOB-PM approach \cite{Damour:2025uka}.

An alternative route to a self-consistent bound orbit model is to use the local-in-time post-Minkowskian Hamiltonian derived by Dlapa \textit{et. al.} in \cite{Dlapa:2024cje, Dlapa:2025biy}. 
This has the advantage that it doesn't employ a velocity expansion, at the cost of an undetermined coefficient for the nonlocal part of the bound dynamics. This approach could be supplemented with PN information to determine the nonlocal coefficient to the desired PN order. As we will be focusing on scattering trajectories for the remainder of this work, it is most appropriate to employ fully dissipative PM information, including the nonlocal tail effects. 

\clearpage

\subsection{Comparison to numerical relativity}
When comparing analytical models to NR, it is often convenient to work with an appropriate set of gauge-invariant observables. 
The scattering angle is a natural gauge invariant observable~\cite{Damour:2014afa}, being gauge invariant at each perturbative order\footnote{Scattering amplitudes, from which PM scattering angles are derived, are gauge-invariant quantities that satisfy the Ward identities, providing a natural consistency check for QFT-based methods~\cite{Mogull:2020sak}.} as well as for the full nonperturbative result. 
As discussed in Sec.~\ref{subsection:eob}, the flexibility in how the EOB-PM Hamiltonians are constructed introduces gauge dependence in the resummed scattering angle~\cite{Damour:2022ybd,Khalil:2022ylj}. 
We demonstrate that this dependence can be particularly strong, with the choice of gauge and resummation strategy playing a crucial role in the accuracy of the model across the parameter space. 

The comparison of analytical models to BBH NR simulations has been explored in a range of studies~\cite{Damour:2014afa,Damour:2022ybd,Hopper:2022rwo,Rettegno:2023ghr,Swain:2024ngs,Albanesi:2024xus,Long:2025nmj}, and recently extended to BNS simulations in~\cite{Fontbute:2025vdv} to explore tidal effects. 
Here we focus on the comparisons to the NR data presented in~\cite{Swain:2024ngs}, focusing on two specific energies, $\Gamma_1 \approx 1.0225 $ and $\Gamma_7 \approx 1.2169$, where the notation has been chosen for consistency with~\cite{Swain:2024ngs}. 
These energies represent the lowest and highest energy simulations in~\cite{Swain:2024ngs}, allowing us to stress-test the different gauge choices across a range of energy scales.  

We analyse multiple approaches in computing the EOB scattering angles. 
Each approach has in common the definition of the scattering angle,
\begin{align} \label{scattering-integral}
    \theta_{\rm EOB} + \pi = - 2\int_{r_{\rm min}}^{\infty} dr\ \frac{\partial p_r}{\partial \pphi}.
\end{align}
Employing the form of the radial momentum defined by the impetus formula Eq.~\eqref{impetus-formula}, this may be expressed as
\begin{align}
    \theta_{\rm EOB} + \pi = -2\pphi  \int_{r_{\rm min}}^{\infty} \frac{dr}{\sqrt{p_{\infty}^2 - \pphi^2 \, r^{-2} + w(r,\gamma, \nu)}}.
\end{align}
How we define the $w$-potential determines the specific resummation of the scattering angle in the EOB-PM model. There are a number of different choices and combinations of choices permitted. For example, the SEOB model of reference \cite{Buonanno:2024vkx} employs Schwarzschild coordinates as the coordinate gauge choice, PS* for the EOB gauge and it does not PM expand the w-potential. Therefore the $w(r,\gamma,\pphi,\nu)$ in the integrand is given by [using these choices in conjunction with Eq.~\eqref{w-potential-definition}]
\begin{widetext}
\begin{align}
w_{\rm SEOB-PM}(r,\pphi,\gamma,\nu) &= \left( 1 - \frac{1}{r^2\left(1 - \frac{2}{r}\right)} \right)\frac{\pphi^2}{r^2} \quad + \left( \frac{1}{\left(1- \frac{2}{r} + \Delta A\right)\left(1-\frac{2}{r}\right)}-1 \right)\gamma^2 \quad - \left(\frac{1}{1-\frac{2}{r}} - 1 \right).
\end{align}
\end{widetext}

On the other hand, the $w_{\rm eob}$ model of reference \cite{Damour:2022ybd} uses isotropic coordinate gauge, PS EOB gauge and chooses to PM expand the $w$-potential, given by the following,
\begin{align}
    w_{\rm eob}(r, \gamma, \nu) = \sum_i \frac{\bar{w}_{i}(\gamma, \nu)}{\bar{r}^i},
\end{align}
where the $\bar{w}_i$ are defined in Eq.~\eqref{w-potential-coefs-ps-isotropic}.

The integrals (\ref{scattering-integral}) are all computed numerically rather than analytically, due to the complex functional form that the resummed $w$-potential takes. 
Despite this, for the PM expanded w-potentials one can compute (at the 4PM order) the integrals analytically, with results in terms of Elliptic functions \cite{Damour:2022ybd}. 
Our numerical results agree with their analytical calculations where applicable.   

\begin{figure*}[t]
        \centering
        \includegraphics[width=\linewidth]{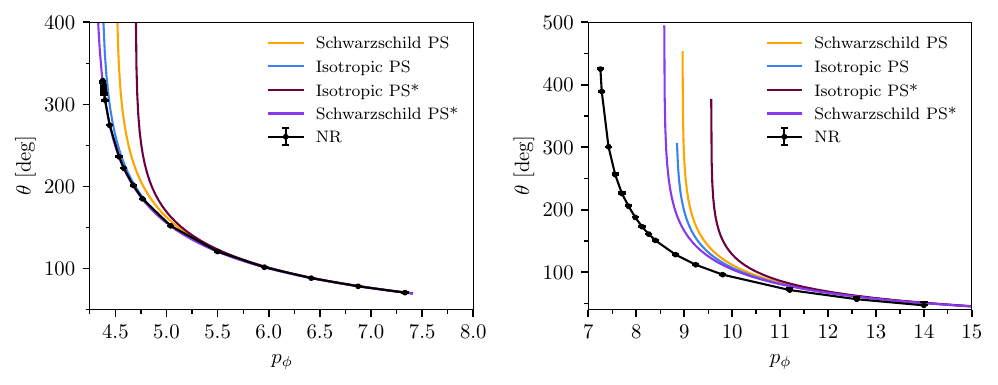}
        \caption{A comparison of nonspinning SEOB-type (resummed potential) models with the numerical relativity data from \cite{Swain:2024ngs}. We vary the coordinate gauge and EOB-gauge for an equal mass nonspinning configuration, at 4PM order. Left: the lower energy comparison at $\Gamma_1\approx 1.02$. Right: the higher energy comparison at $\Gamma_7 \approx 1.22$. The curves are truncated when the model predicts a plunge.
        \label{fig:scattering-angle-both-energies}
        }
\end{figure*}

Starting with the resummed $w$-potentials, Fig.~\ref{fig:scattering-angle-both-energies} shows that varying the coordinate gauge and EOB gauge has a large impact on the convergence of the model to numerical relativity. 
In particular, it shows that there is a pairing up of EOB gauge and coordinate gauge, with Schwarzschild coordinates performing far better in the PS* gauge as opposed to PS. 
The situation for isotropic coordinates is the opposite, and these behaviors align with the discussion in the precious section regarding the effective potentials.
Fig.~\ref{fig:schwarzschild_high_E_comparison} shows the high energy comparison using Schwarzschild coordinates, highlighting the impact of the EOB-gauge on the final model. 
In Schwarzschild coordinates, PS gauge performs significantly worse than PS*, agreeing with the results of \cite{Khalil:2022ylj}. 
It is important to note that in that particular paper, the EOB gauge is only discussed for models using Schwarzschild coordinates, missing the subtlety of the interplay between coordinate and EOB gauge choice. Figure \ref{fig:angle-comparison-PS-both-energies} highlights the poor
performance of Schwarzschild coordinates in the PS gauge.

\begin{figure*}[t]
    \centering
    \includegraphics[width=\linewidth]{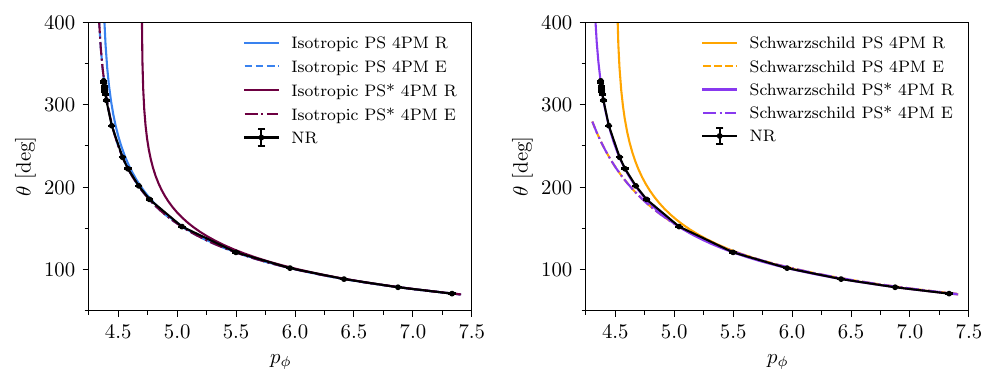}
    \caption{Comparing the PM expanded (E) and resummed (R) potential models across the four gauge choices, comparing at the lower energy $\Gamma_1$ for an equal mass system.}
    \label{fig:Expanded-vs-resummed}
\end{figure*}

Another interesting feature of these models, as illustrated in Fig.~\ref{fig:Expanded-vs-resummed}, is the differing impact of PM expanding the $w$-potential across the gauge choices. In particular, with certain choices, a PM expansion can actually \textit{improve} the convergence to NR. In particular, if we consider the isotropic coordinate model in a PS gauge, it performs reasonably well with the resummed $w$-potential, but still worse than the model with Schwarzschild PS*. However, if we PM expand the $w$-potential, it vastly improves NR agreement, particularly at the higher energy. 
The comparisons in Fig.~\ref{fig:Expanded-vs-resummed} also provide a consistency check that, in the PM expanded regime, the two EOB gauges are equivalent.

Through comparing the scattering angles to numerical relativity data, we have tested all of the different combinations of gauges and $w$-potential definition. 
We find that the best performing gauge to be Schwarzschild coordinates combined with a PS* EOB gauge, which aligns with the SEOB-PM model of Ref.~\cite{Buonanno:2024vkx}. 
A close second is isotropic radial coordinates combined with either EOB gauge, with the caveat that the $w$-potential must be PM expanded. 
The latter is equivalent to the $w_{\rm eob}$ model of Ref.~\cite{Damour:2022ybd}. 
It is worth noting that for the PM expanded $w$-potentials, the choice of PS or PS* in the initial construction does not influence the final model, consistent with the equivalence of the potentials.

\subsubsection{Introducing 5PM information}

The formalism used in the previous section can be readily extended to include the new radiative 5PM-1SF results of Ref.~\cite{Driesse:2024feo}. We restrict our attention to the two best performing models, namely, Schwarzschild coordinates with a PS* EOB gauge and isotropic coordinates with a PS EOB gauge. We further PM expand the radial momentum of the latter model due to the improvement found in the previous section. We note here that these models correspond to the SEOB-PM and $w_{\rm eob}$ models.

For the SEOB-PM type model, we employ the radial momentum of Eq.~\eqref{radial-momentum-PS*}, with deformation coefficients included in the $A$-potential, now with an extra $\alpha_{(5)}$-term. For the $w_{\rm eob}$ type model we employ the radial momentum of Eq.~\eqref{prsq-PM}, with an isotropic radial coordinate and include a $w_5$ term. The $w_5$ coefficient can be related to the $q_i$ deformation coefficients by 
\begin{align}
    \bar{w}_5(\pphi, \gamma, \nu) = 6\gamma^2 - \frac{\bar{q}_2}{2} - \frac{3\bar{q}_3}{2} - 2q_4 - q_5.
\end{align}
Alternatively one can simply treat the $w_i$ as the fundamental deformation functions of the model and match these to the scattering angle, see Appendix ~\ref{appendix: w-potentials} for further details and the key relations.

\begin{figure*}[t]
    \centering
    \includegraphics[width=\linewidth]{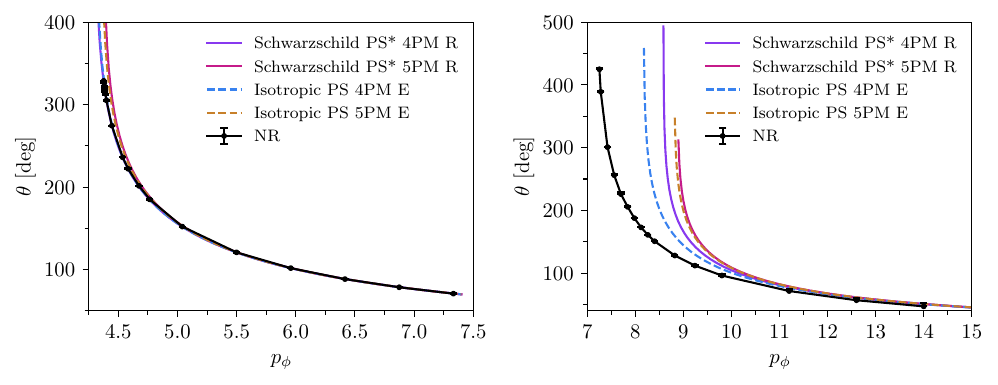}
    \caption{Plots showing the performance of the 5PM-1SF information across the two best combinations of gauge and $w$-potential definition. The left panel shows an equal mass configuration at the lower energy $\Gamma_1$, with the right plot at the higher energy $\Gamma_7$. We see that in both cases the $5$PM-$1$SF information is slightly worse than the corresponding 4PM results. As in previous plots, the curves either asymptote or terminate when the model predicts a plunge.}
    \label{fig:5PM-comparisons}
\end{figure*}

We include the 5PM-1SF deformation coefficients in an ancillary \textit{Mathematica} file. 
It is worth noting that the complexity of the expressions increases dramatically from 4PM to 5PM, which may adversely affect the incorporation of these terms into bound-orbit models, particularly due to the computational overhead associated with the size of the expressions and the presence of special functions. 
We can see from Fig.~\ref{fig:5PM-comparisons} that the 5PM information slightly worsens the agreement with NR, at both the lower and higher energy scales.
It is worth emphasizing that the 5PM orbital sector is as yet incomplete, so these preliminary results are only a comment on the currently available analytic information.
In terms of gauge choices, it is generally consistent with the 4PM analysis, but with the $w_{\rm eob}$ framework slightly outperforming the SEOB-PM type model. 
Given that the changes are minimal, the discussion on the gauge choices at 4PM still holds at the 5PM level.

\subsubsection{Pad\'e resummation}

As discussed in~\cite{Damour:2022ybd} and~\cite{Swain:2024ngs}, Pad\'e resummation of the $w$-potentials can be used to improve the agreement with NR. 
In particular, \cite{Swain:2024ngs} found that at $5 \rm PM$ a $P^4_1$ Pad\'e had the best agreement with numerical relativity, with a calibrated 5PM parameter. Now that the full $5$PM-$1$SF information is available, we can test the Pad\'e resummation on the purely analytical information. 

A $P^{n}_{m}$ Pad\'e approximant is a resummation of a $k^{th}$ order polynomial, say $f(x)= a_0 + a_1 x + a_2 x^2 + ... + a_k x^k$, as a rational function 
\begin{align}
    P^{n}_{m}(x) = \frac{N_n(x)}{M_m(x)}, 
\end{align}
with the conditions that $n + m =k$ and 
\begin{align}
    T_k [P^{n}_{m}(x)] = f(x),
\end{align}
where $T_k[\cdot]$ is the Taylor expansion operator. 

Pad\'e resummation has long been used to improve the behavior of the potentials in the EOB formalism, particularly for PN \cite{Damour:1997ub, Damour:2000we}, as it can help control the divergent behavior of the perturbative information in certain regions of parameter space.
More recently, it has been used in the LEOB approach to bound orbits, where a $(4,1)$ Pad\'e was used to resum the PM information incorporated in the $A$-potential \cite{Damour:2025uka}. 

Here, we are interested in applying Pad\'e resummation to the PS* Schwarzschild gauge, with the resummed definition of the $w$-potential (SEOB-PM model) and also to the PS/PS* isotropic gauge with a PM expanded $w$-potential ($w_{\rm eob}$ model). 
Due to the different structure of each approach, we apply the resummation in different ways. For the former we directly resum the $A$-potential, which includes the deformation coefficients. For the latter we will resum the PM expanded $w$-potential. In both cases we aim to improve the behavior of the PM information incorporated into the model. 

A further condition on the Pad\'e approximants for the SEOB-PM model will be that they recover the correct $\nu \rightarrow 0$ limit. This is less restrictive for the $w_{\rm eob}$ model, which does not recover the nonperturbative test-mass limit in any case. 
The Pad\'e orders that we consider all respect the test mass limit such that as $\nu \rightarrow 0$, we have that $A \rightarrow 1 - 2/r$. 
In particular, $P^{2}_2$, $P^{1}_{3}$ and $P^{3}_{1}$ at 4PM and $P^{2}_{3}$, $P^{3}_{2}$, $P^{4}_{1}$ and $P^{1}_{4}$ at 5PM. 

\begin{figure}
    \centering
    \includegraphics[width=\linewidth]{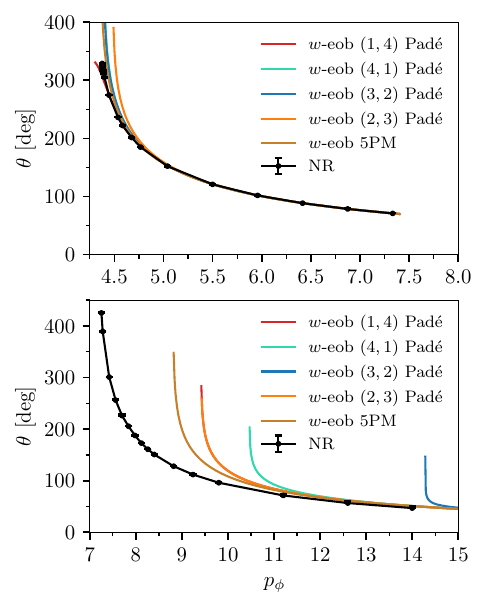}
    \caption{A comparison of the different Pad\'e approximants considered in the $w$-eob model at $5$PM, the upper plot shows the lower energy comparison $\Gamma_1$ and higher energy comparison $\Gamma_7$. As in previous plots, the lines terminate when the model predicts a plunge.}
    \label{fig:Pade-weob-main-text}
\end{figure}

At 4PM, SEOB-PM benefits from resummation of the $A$-potential, with a $(2,2)$ Pad\'{e} approximant significantly improving agreement with NR at higher energies. 
For $w_{\rm eob}$, we found little-to-no improvement with any Pad\'e resummation, with the vanilla $w_{\rm eob}$ model being the most robust across the energies tested. 
This highlights that while Pad\'{e} resummation can be beneficial in certain contexts, it does not universally provide a robust strategy for resumming higher-order PM information, at least in the scattering regime. 
A dedicated study of how this impacts the bound orbit dynamics, examining both the $A$ and $w$-potentials, would be required to assess the efficacy of Pad\'{e} resummation in that context. 
This is a nontrivial statement given the different analytic structure of the potentials when conservative local contributions and additional PN information are included~\cite{Damour:2025uka}.
At $5$PM we find a similar conclusion, at the lower energy scale there is a negligible difference between the Pad\'e approximants and the original model but at the higher energy scale there are much larger discrepancies, indicated in Fig.~\ref{fig:Pade-weob-main-text}. In particular we find a $(3,2)$ Pad\'e very slightly increases the performance at higher energies for the SEOB-PM model, but for the $w_{\rm eob}$ model all of the Pad\'e approximants perform significantly worse than the standard 5PM $w_{\rm eob}$. Figures \ref{fig:SEOB-Pade-comparisons} and \ref{fig:weob-pade-comparisons} in App.~\ref{appendix-additional-plots} contain more comparisons across the different models considered.

It is interesting to note that our results differ from those of Ref.~\cite{Swain:2024ngs}, where a $5$PM parameter was calibrated to numerical relativity. 
In that work the authors found that a Pad\'e approximation could improve the agreement of the $w$-potential to a potential extracted from the NR simulation quite considerably. However, our results are purely based on analytical information, rather than with a numerical NR tuned parameter, which is likely the reason for the discrepancy.

\section{Incorporating Spin}\label{section:Spinning-Sector}

We now turn to the spinning case. The procedure for computing the EOB metric is the same as the nonspinning case, but now using a test mass in a Kerr metric as the basis for the model \cite{Damour:2001tu}. The effective Hamiltonian takes the following form
\begin{align}
    H_{\rm eff} = H_{\rm SO} + H_{\rm orb}, 
\end{align}
where the orbital part can be written in terms of deformed metric potentials
\begin{align}
    H_{\rm orb} = \sqrt{A_{\rm eff} \left(1 + \frac{\pphi^2}{r^2} + (1 + B_{\rm np})p_r^2 + B_{\rm npa}\frac{\pphi^2 \chi_+^2}{r^2}\right )},
\end{align}
where $\chi_+ = a_1 + a_2$ replaces the Kerr parameter in Eq.~\eqref{Kerr-hamiltonian}.
The spin orbit term may be written in terms of gyrogravitomagnetic couplings, as in \cite{Damour:2008qf}, in the following way
\begin{align}
    H_{\rm SO} = \pphi(G_{S}S + G_{S_*}S_*).
\end{align} 

There are a number of choices for spin variables used in the literature. 
For consistency with the recent SEOB-PM work~\cite{Buonanno:2024vkx}, we use the variables $\chi_\pm = a_\pm/(GM)$. It is common to see the variables $S= S_1 + S_2 = m_1 a_1 + m_2 a_2$ and $S_* = (m_2/m_1)S_1 + (m_1/m_2)S_2 = m_2a_1 + m_1 a_2$ used in the literature, they can be further related to $a_+$ and $a_-$ by
\begin{align} 
\label{s-a-relations}
    S &= \frac{M}{2}(a_+ + \delta a_-), \\
    S_* &= \frac{M}{2}(a_+ - \delta a_-),
\end{align}
where $M = m_1 + m_2$ is the total mass of the system and $\delta = (m_1 - m_2) / 2$. Finally, our variables of choice $\chi_\pm$ are related through $S = 1/2 (\chi_+ + \delta \chi_-)$ and the analogous relation for $S_*$.

It is worth noting here that in the spinning sector we employ a physical PM power counting scheme~\cite{Rettegno:2023ghr,Buonanno:2024vkx}, which takes into account the scaling of the spins with $G$, i.e. $a \leq G M$. 
Each appearance of $a$ acts to increases the PM order of the expression. 
This is in contrast to the formal PM power counting, in which the $(l + s + 1)$PM order corresponds to loops of order $l$ and spins of order of $s$, see~\cite{Rettegno:2023ghr,Buonanno:2024vkx}.  
Another subtlety when dealing with spinning PM results is the distinction between the covariant angular momentum, which is commonly used in scattering amplitudes based calculations and canonical angular momentum, which is the definition appropriate for comparison to NR. One can convert between the two using the relation (valid in the aligned spin limit),
\begin{align}
    \pphi^{\rm (can)} = \pphi^{\rm (cov)} - \frac{\Gamma - 1}{2 \nu} \left( \chi_+ - \frac{\delta}{\Gamma} \chi_- \right).
\end{align}
In this section we always work with the canonical definition of angular momentum, so we will drop the explicit label.

\subsection{Spinning EOB-PM models}

\subsubsection{SEOB-PM}
As was discussed in Sec.~\ref{subsection:seob+weob}, the SEOB-PM model~\cite{Buonanno:2024vkx} is based on the Kerr Hamiltonian expressed in Boyer-Lindquist coordinates. 
The SEOB-PM model adopts the PS$\ast$ gauge, incorporating PM information via deformations to the metric potentials and gyrogravitomagnetic spin couplings. 
A key choice is that even-in-spin corrections are incorporated into the $A$-potential, 
\begin{align} 
\label{seob-delta-A}
    A_{\rm eff}(r,\gamma,a_\pm) &= \frac{1 - \frac{2}{r} + \frac{\chi_+^2}{r^2} + \sum_{n} \frac{\Delta A^{(n)}}{r^n}}{1 + \frac{\chi_+^2}{r^2}(1 + \frac{2}{r})},  \\
    \Delta A^{(n)} &= \sum_{s=0}^{\lfloor \frac{n-1}{2}\rfloor} \sum_{i=0}^{2s}\alpha_{(2s-i, i)}\delta^{\sigma(i)} \chi_+^{2s-i} \chi_-^{i},
\label{seob-delta-Aexp}
\end{align}
where 
\begin{align}
    \sigma(n) = \begin{cases}
    0, \qquad n\ \rm even, \\
    1, \qquad n\ \rm odd,
    \end{cases}
\end{align}
\noindent
and odd-in-spin terms into the gyrogravitomagentic couplings,
\begin{align}
    g_{\chi_{\pm}} &= r^2 \sum_{n\geq 2} \frac{\Delta g_{a_{\pm}}^{(n)}}{r^n} \\
    \Delta g_{\chi_{+}}^{(n)} &= \sum_{s=0}^{\lfloor \frac{n-2}{2}\rfloor} \sum_{i=0}^s \alpha^{(n)}_{(2(s-i)+1, 2i)} \chi_+^{2(s-i)}\chi_-^{2i}, \\
    \Delta g_{\chi_{-}}^{(n)} &= \sum_{s=0}^{\lfloor \frac{n-2}{2}\rfloor} \sum_{i=0}^s \alpha^{(n)}_{(2(s-i), 2i+1)} \chi_+^{2(s-i)}\chi_-^{2i}.
\end{align}
The effective Hamiltonian is then given by
\begin{align}
    &H_{\rm eff} = \frac{\pphi(g_{\chi_+}\chi_+ + g_{\chi_-}\delta \chi_-)}{r^3 + a_+^2 (r + 2)}  \\
    &+ \sqrt{A_{\rm eff} \Big(1 + \frac{\pphi^2}{r^2} + (1 + \Bnp)p_r^2 + \Bnpa \frac{\pphi^2 \chi_+^2 }{r^2} \Big)}, \notag
\end{align}
where the Kerr spin is taken to be $a_+ = M \chi_+$ and $p_{\phi}$ corresponds to the orbital angular momentum $L$ of the effective test-body. 
The gyrogravitomagnetic factors are expressed in a gauge in which $g_{\chi_+}$ and $g_{\chi_-}$ depend on $1/r$ and $L^2 / r^2$ but not on $p^2_r$~\cite{Khalil:2023kep,Buonanno:2024vkx}, in contrast to other gauges that could be employed~\cite{Damour:2008qf,Barausse:2009xi,Nagar:2011fx,Placidi:2024yld}.
The $\alpha$ coefficients are calculated using the methods outlined in the non-spinning sector, and our coefficients are in agreement with those presented in~\cite{Buonanno:2024vkx}. 

\subsubsection{Spinning \texorpdfstring{$w_{\rm eob}$}{w-eob} formalism}
Within the $w_{\rm eob}$ framework, we can construct a spinning model following the expansion analogous to that presented in~\cite{Rettegno:2023ghr}. 
A key difference here is that Ref.~\cite{Rettegno:2023ghr} employed a formal PM counting scheme whereas we use the physical power counting scheme. 
Recall that in the non-spinning sector, the $w_{\rm{eob}}$ model defined a PM-expansion of radial momentum of the form
\begin{align} \label{mass-shell-spinning-weob}
    p_r^2 &= (\gamma^2 - 1) - \frac{p_\phi^2}{r^2} + \sum_{n\geq1} \frac{w_n(\gamma, \nu)}{r^n}, 
\end{align}
where the PM deformations are systematically incorporated through a calibration procedure that matches the scattering angle, expressed in terms of the undetermined $w$-potential coefficients $w_n(\gamma, \nu)$, to the analytically known PM scattering angles.
Spins can then be introduced by performing a spin-expansion~\cite{Rettegno:2023ghr}, such that 
\begin{align} \label{spin-expansion-weob}
    w_n(\pphi, \gamma, \nu, \chi_\pm) &= w^{(n)}_{(0,0)}(\gamma, \nu)  + w^{(n)}_{\rm SO}(\pphi, \gamma, \nu, \chi_\pm) \\
    & \notag \qquad + w^{(n)}_{\rm SS}(\pphi, \gamma, \nu, \chi_\pm).
\end{align}
The $w^{(n)}_{(0,0)}$ term corresponds to the potentials in the orbital sector, $w^{(n)}_{\rm SO}$ to the spin-orbit (SO) terms capturing the odd-in-spin deformations, and $w^{(n)}_{\rm SS}$ to the spin-spin (SS) coefficients that incorporate the even-in-spin deformations. 
To maintain consistency with the physical PM counting scheme, each term $w_n$ is expanded to order $(n-1)$ in the spins.
The spin-orbit part of the potential takes the following form,
\begin{align}
    w^{(n)}_{\rm SO}(r, \pphi, \gamma, \nu, \chi_\pm) = \frac{\pphi}{r^3} \left( w_{\chi_+} \chi_+ + w_{\chi_-}\delta \chi_- \right),
\end{align}
where 
\begin{align}
    w_{\chi_+} &= w^{(2)}_{(1,0)} + \frac{w^{(3)}_{(1,0)}}{r} \\& \qquad + \frac{w^{(4)}_{(1,0)} + w^{(4)}_{(2,1)}\chi_+ \chi_- + w^{(4)}_{(3,0)}\chi_+^2}{r^2} +... \nonumber
\end{align}
and a similar expression holds for $w_{\chi_-}$.
$w_{\chi_-}$ has a similar expression, but with $ + \rightarrow -$:
\begin{align}
    w_{\chi_-} &= w^{(2)}_{(0,1)} + \frac{w^{(3)}_{(0,1)}}{r} + \\& \qquad \frac{w^{(4)}_{(0,1)} + w^{(4)}_{(1,2)}\chi_+ \chi_- + w^{(4)}_{(0,3)}\chi_-^2}{r^2}+... \nonumber
\end{align}
\noindent
The spin-spin potential has the following functional form
\begin{align}
    w_{\rm SS}^{(n)} = \sum_{s=0}^{\lfloor \frac{n-1}{2}\rfloor} \sum_{i=0}^{2s}w^{(n)}_{(2s-i, i)}\delta^{\sigma(i)} \chi_+^{2s-i}\chi_{-}^i .
\end{align}

The structure of the $w_{\text{SO}}$ potential mirrors that of the spin-orbit components in the effective Hamiltonians presented above. 
Since the $w_{\rm{eob}}$ model can be recovered through PM expansions of SEOB-PM-type models, we construct explicit mapping relations between the $\alpha$ coefficients of the previous models and the $w$-potential coefficients, analogous to the procedure in the non-spinning sector. 
The explicit expressions are given in Appendix ~\ref{appendix: w-potentials}.

Employing these potentials, substituting Eq.~\eqref{spin-expansion-weob} into Eq.~\eqref{mass-shell-spinning-weob} defined a radial momentum that we can use to compute spinning $w$-effective-potential coefficients. 
\begin{align}
    p_r^2 = (\gamma^2 - 1) - \frac{\pphi^2}{r^2} + \sum_{n\geq1} \frac{w_n(\pphi,\gamma,\nu,S_1, S_2)}{r^n}.
\end{align}
We note that $n$ counts the physical PM order, taking into account the factors of $G$ coming from the spins. 

With our expansion of the $w$-potentials we can see that one can construct a one-way mapping between the radial momentum employed in the SEOB-PM model and the $w_{\rm eob}$ models, by PM expanding out each part. The gyrogravitomagnetic couplings expand out to $w_{\rm SO}$ and the orbital part of the hamiltonian maps onto $w_n$ and $w_{\rm SS}$, the explicit maps between the $\alpha$ coefficients and the $w$-potential coefficients may be found in Appendix ~\ref{appendix: w-potentials}.

\subsubsection{Centrifugal radius}
As outlined in Sec.~\ref{subsection:kerr-metric}, the centrifugal radius reveals an elegant structure to the Kerr $A$-potential, showcasing how it can be factorized into a Schwarzschild-like term and a multiplicative correction factor that encodes the multipolar structure of Kerr~\cite{Damour:2014sva}.
The centrifugal radius was defined as
\begin{align} \label{centrifugal-rad}
    r_c^2 &= r^2 + a^2 + \frac{2Ma^2}{r}, 
\end{align}
where $r_c$ was been rescaled in the usual way, $r_c = R_c / (GM)$. 
Using this radius in the Kerr Hamiltonian incorporates spin-spin effects while casting the functional form into something reminiscent of the Schwarzschild Hamiltonian,
\begin{align}
    H_{\rm Kerr, eq}^{\rm orb} = \sqrt{A^{\rm eq}(r) \left( 1 + \frac{\pphi^2}{r_c^2} + \frac{p_r^2}{B^{\rm eq}(r)} \right) },
\end{align}
as outlined in Sec.~\ref{subsection:kerr-metric}. 
The odd-in-spin part of the Hamiltonian takes the form
\begin{align}
    H_{\rm SO} &= \pphi(G_S S + G_{S_*} S_*), \\
    G_i &= \frac{g_i}{r r_c^2}, \\
    H_{\rm SO} &= \frac{\pphi}{rr_c^2} \left( g_{S}S + g_{S_*}S_* \right).
\end{align}
The leading-order values of the spin-orbit couplings $g_S$ and $g_{S_*}$ are determined by the test-mass limit, such that
\begin{align}
    g_S^{(\rm LO)} = 2, \qquad
    g_{S_*}^{(\rm LO)} = \frac{3}{2}.
\end{align}
We can transform these gyrogravitomagnetic terms into the same functional form as employed in SEOB-PM model using Eqs.~\eqref{s-a-relations}, 
\begin{align}
    g_{a_+} = \frac{1}{2}(g_S + g_{S_*}), \\
    g_{a_-} = \frac{1}{2}(g_S - g_{S_*}).
\end{align}
The leading-order values of the coupling functions are then given by
\begin{align}
    g_{a_+} = \frac{7}{4}, \qquad g_{a_-} = \frac{1}{4}.
\end{align}

In the PM expansion, we need to track the order of $r_c$ carefully due to its spin dependence. 
One way to handle this is to treat $r_c$ in the same way as the other metric potentials and express it as a perturbative PM expansion
\begin{align}\label{centrifugal-rad-pm}
    [r_c(\chi,r)]^2 = r^2 \left[ 1 + \sum_{i \geq 2} \frac{r_{c,i}(\chi, \gamma)}{r^i} \right].
\end{align}
At leading order, $r_c^2$ is $\mathcal{O}(G^{-2})$ so the PM expansion within the square brackets corresponds to the physical PM order.
We will explore two strategies for incorporating PM information into the Hamiltonians based on the centrifugal radius. 
First, we consider a gauge in which $r_c$ is fixed to the Kerr limit and incorporate PM information by deforming the Schwarzschild-like term in the $A$-potential
\begin{align}
    A^{\rm eq}_{\rm eff}(r, \gamma, \nu) = \left[ 1 - \frac{2}{r_c} + \Delta A \right] \frac{ \Big(1 + \frac{2}{r_c} \Big)}{ \Big( 1 + \frac{2}{r} \Big)} ,
\end{align}
where $\Delta A$ corresponds to Eq.~\eqref{seob-delta-Aexp}. 
We denote this model as the centrifugal-Kerr model given that the centrifugal radius is fixed to the Kerr limit.

The second approach incorporates only the nonspinning deformations into the $A$-potential, analogous to the PS* prescription, while spin-spin contributions are included via a deformation of the centrifugal radius $r_c$.
The nonspinning deformations to the $A$-potential take the form
\begin{align}
    \Delta\tilde{A}(r_c) = \sum_{i \geq 2} \frac{\tilde{\alpha}^{(i)}_{(0,0)}}{\tilde{r}_c^i} \,,
\end{align}
where the tilde notation is used to distinguish the coefficients and the deformations to the metric potentials from their counterparts defined in Eq.~\eqref{seob-delta-Aexp}.
The resulting deformation to the centrifugal radius can be expressed as 
\begin{align}
    \tilde{r}_c^2(r, \chi_+, \chi_-, \gamma, \nu) = r^2 \left( 1 + \frac{\chi_+^2}{r^2} \left(1 + \frac{2}{r} \right) + \Delta \chi \right),
\end{align}
where we note that $\chi_{\pm}$ is dimensionless by construction and therefore does not contain any factor of $G$. 
Note that $\Delta \chi$ starts at the physical 3PM order, leading to the following PM expansion
\begin{align}
    \Delta \chi  = \sum_{n \geq 3}\sum_{s=0}^{\lfloor \frac{n-1}{2} \rfloor} \sum_{i=1}^{2s} \frac{1}{r^n}  \tilde{\alpha}^{(n)}_{(2s-1, i)} \delta^{\sigma(i)} \chi_+^{2s-1}\chi_-^i.
\end{align}
Since we are including PM deformations within the centrifugal radius, we denote this second variant as centrifugal-PM. 

As we shall see in the next section, the two models exhibit varying levels of accuracy. The centrifugal-Kerr model performs comparably to the SEOB-PM model, while the centrifugal-PM model shows better performance at large positive and negative spins but suffers a degradation in accuracy at smaller spins.

\subsubsection{Comparison to numerical relativity}
We compare the various models with the numerical relativity data reported in Table II of~\cite{Rettegno:2023ghr}, which corresponds to an equal-mass, equal-spin series with fixed energy and angular momentum ($\pphi \approx 4.5$, $\Gamma \approx 1.02$) but with varying equal-spin magnitudes $\chi_+$.
As outlined above, there are three main classes of models in the spinning sector: SEOB-PM, $w_{\rm eob}$, and models based on the centrifugal radius. We will also consider variants of these models. 
In particular, we find considerable improvement when setting the even spin-spin coefficients to zero in the SEOB-PM and centrifugal-Kerr models, retaining only the spin dependence in the gravitomagnetic factors.

\begin{figure*}[]
    \centering
    \includegraphics[width=\linewidth]{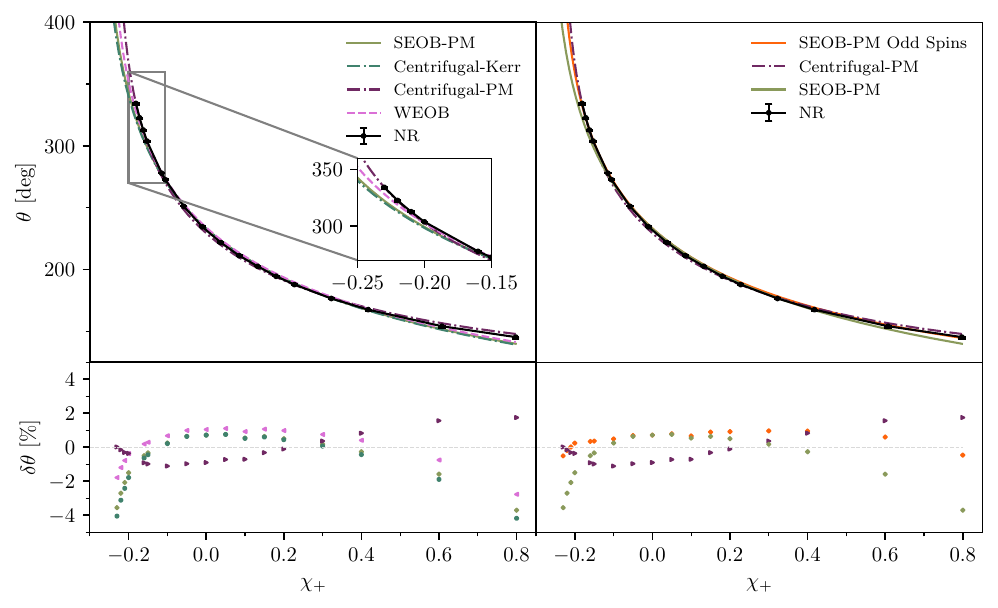}
    \caption{Comparisons of the different spinning EOB-PM models against NR. Both plots show an equal mass configuration with aligned and equal spins, varying the spin magnitude. Left upper: comparing all of the models which incorporate all the available spinning PM information. Lower: a pointwise percentage comparison to the NR data points. Right: comparing the centrifugal-PM model with the SEOB-PM odd spin variant.
    \label{fig:spinning-nr-delta}
    }
\end{figure*}

In Fig.~\ref{fig:spinning-nr-delta}, we compare the performance of the four models defined above to the NR scattering angles. 
Each model demonstrates good agreement (within $2\%$) for small positive spins, but starts to deviate for larger positive and negative spin magnitudes.
Both the SEOB-PM and centrifugal-Kerr models are too repulsive at negative spins, resulting in a smaller scattering angle than predicted by NR at the $\sim 4\%$ level. The $w_{\rm eob}$ model performs marginally better but is still slightly too repulsive. 
This is potentially attributable to the PM-expanded form of the $w$-potential moderating the behavior at negative spins.
The centrifugal-PM model, however, shows the best agreement with the NR data, maintaining errors that are below $2\%$ across all data points tested. 
This suggests that the centrifugal radius formulation offers a promising method for resumming spinning PM information within the EOB-PM framework.

We have also compared variants of the SEOB-PM and centrifugal-Kerr, in which we simply switch off the spin-spin information contained in the $A$-potential. This, although seemingly an ad-hoc prescription greatly improves the agreement to NR across the data points tested, as shown in the right panel of Fig.~\ref{fig:spinning-nr-delta}. The SEOB-PM model with even spin-spin information removed maintains a sub $1\%$ deviation across these points, suggesting that the spin-spin information in the $A$-potential weakens the SEOB-PM model. This could suggest that the Kerr-like description of the spin-spin interactions could be more robust than the PM even-in-spin-information, however more investigation is required to fully understand this discrepancy.
Figures 15–17 provide additional comparisons across the different models, high-lighting the impact of incorporating even-in-spin terms.
More numerical simulations would be required to robustly test this prescription across more of the parameter space.

\section{Conclusions}

In this work we have systematically tested the impact of gauge choices on PM based EOB models, building on earlier works  \cite{Buonanno:2024vkx, Damour:2022ybd, Rettegno:2023ghr, Khalil:2022ylj}. By constructing from the ground-up a variety of models in the spirit of the SEOB-PM model \cite{Buonanno:2024vkx} and building the related $w_{\rm eob}$ type models \cite{Damour:2022ybd, Rettegno:2023ghr}, we have found that in the nonspinning limit there is a delicate relationship between coordinate gauge, EOB gauge and PM expansion that can have a drastic impact on the performance of the model, as seen in Fig~\ref{fig:Expanded-vs-resummed}. The gauge corresponding to the SEOB-PM model (Schwarzschild PS*) performed the best when paired with the resummed $w$-potential. A close second was the isotropic coordinate gauge, paired with a PM expanded potential (in which case the EOB gauges are equivalent) corresponding to the $w_{\rm eob}$ model, with all other choices performing worse when compared with NR. We have highlighted that the two gauge choices, coordinate and EOB, seem to be interrelated. In particular, Schwarzschild coordinates must be used with the PS* gauge whereas isotropic requires PS.

We also incorporated the recent dissipative $5$PM-$1$SF information from \cite{Driesse:2024feo} and found that the agreement worsens in comparison to $4\rm PM$. This may be due to the incomplete nature of the $5$PM sector, motivating the need for its completion, allowing tests of the true impact of higher perturbative orders on EOB models. The 5PM information does not drastically change the conclusions regarding gauge choices, however it does show a slight preference toward the PM expanded isotropic coordinate model ($w_{\rm eob}$) over the SEOB-PM gauge.

As discussed in \cite{Damour:2022ybd, Swain:2024ngs}, the use of Pad\'e approximants to improve the convergence of the models to NR has been explored, but with limited success.
In particular, Pad\'e resummation was not found to be universally robust but were able to demonstrate significant improvements at high energies, in agreement with~\cite{Swain:2024ngs}, which can be seen in Fig~\ref{fig:SEOB-Pade-comparisons} in Appendix \ref{appendix-additional-plots}. 
In constructing the Pad\'e approximants, care was taken to enforce the test-mass limit and to ensure that the potentials reduce to the appropriate expressions in this regime.

In the spinning case we analyzed the existing models, SEOB-PM and $w_{\rm eob}$ and found that NR agreement can be improved significantly by simply switching off the even spin-spin information, suggesting that the way spin-spin information is being incorporated is spoiling the accuracy of the models. 
To that end, we have also constructed two new models based on the concept of the centrifugal radius, one based closely on the SEOB-PM model, centrifugal-Kerr, and another which utilizes the centrifugal radius to incorporate spin-spin information centrifugal-PM. The former has very similar, (albeit slightly worse) accuracy compared with SEOB-PM, but the latter improves NR agreement, significantly for large and negative spins (Fig~\ref{fig:spinning-nr-delta}). 

There are many future directions for this work: To start it would be interesting to see if the analysis of the different gauge choices smoothly transitions to the bound case or if this changes the structure, particularly in the case of higher order PM information. Alongside this comes incorporating information from the recent work by Dlapa \textit{et. al.} \cite{Dlapa:2025biy}, in which a a local conservative Hamiltonian at the fifth PM order was derived. Armed with these results, waveform modeling with the post-Minkowskian expansion can be explored, which would build on the recent works in \cite{Buonanno:2024byg, Damour:2025uka}. The work presented here may be used to inform the choices that are taken in constructing these models, highlighting the care that must be taken with gauge choices in EOB models. Another interesting route would be to build on our centrifugal radius model and extend to higher PM orders, once they are available, also including new physics such as spin precession or tidal effects. This has the potential to improve the modeling of binary systems on generic orbits, particularly in the strong field regime, which could greatly benefit analysis of gravitational wave signals for future high SNR detectors.  

\begin{acknowledgments}
A.C. and G.P. gratefully acknowledge the support of a Royal Society University Research Fellowship, URF{\textbackslash}R1{\textbackslash}221500 and RF{\textbackslash}ERE{\textbackslash}221015. 
A.C. and G.P. would like to thank the Institute for Advanced Studies for hospitality, where parts of this work were completed. 
GP also acknowledges support from a UK Space Agency grant ST/Y004922/1. 
AC would like to thank Shaun Swain for helpful comments. 
\end{acknowledgments}

\appendix

\begin{widetext}

\section{Nonspinning Deformation Coefficients}\label{appendix-deformation-coefs-no-spin}

Here we report the scattering angle coefficients to 4PM order and deformation coefficients up to the 3PM order, for each of the nonspinning gauge choices. The 4PM coefficients may be found in an ancillary file. 

First, the scattering angles in terms of the EOB deformation coefficients are given below.

PS isotropic:
\begin{align}
    \chi_1 &= \frac{2\gamma^2 - 1}{\sqrt{\gamma^2 - 1}} \\
    \chi_2 &= \frac{\pi(15\gamma^2 - 3 - 2\bar{q}_2)}{8} \\
    \chi_3 &= \frac{1}{3 (\gamma^2 - 1)^{\frac{3}{2}}} \Big( 64\gamma^6 - 5 - 9\bar{q}_2 - 3\bar{q}_3 - 3\gamma^4(40 + 4\bar{q}_2 + \bar{q}_3) + 3\gamma^2(20 + 7\bar{q}_2) + 2\bar{q}_3 \Big) \\
    \chi_4 &= \frac{3\pi}{128}(35 + 1155\gamma^4 + 112\bar{q}_2 + 8\bar{q}_2^2 + 64 \bar{q}_3 + 16\bar{q}_4 - 2\gamma^2(315 + 136\bar{q}_2 + 48\bar{q}_3 + 8\bar{q}_4))
\end{align}

PS Schwarzschild:
\begin{align}
    \chi_1 &= \frac{2\gamma^2 - 1}{\sqrt{\gamma^2 - 1}} \\
    \chi_2 &= \frac{\pi}{8}(-3 + 15\gamma^2 - 2q_2) \\
    \chi_3 &= \frac{1}{3(\gamma^2 - 1)^{\frac{3}{2}}}(-5 + 64\gamma^6 - 3q_2 - 3q_3 - 3\gamma^4(40 + 2q_2 + q_3) + \gamma^2 (60 + 9q_2 + 6q_3)) \\
    \chi_4 &= \frac{3\pi}{128}(35 + 1155\gamma^4 + 24 q_2 + 8q_2^2 + 16 q_3 + 16 q_4 - 2\gamma^2(315 + 60q_2 + 24 q_3 + 8q_4))
\end{align}

PS* isotropic:
\begin{align}
    \chi_1 &= \frac{2\gamma^2 - 1}{\sqrt{\gamma^2 - 1}} \\
    \chi_2 &= \frac{-\pi}{8}(3 + \gamma^2(2\bar{\alpha} - 15)) \\
    \chi_3 &= -\frac{1}{3(\gamma^2 - 1)^{\frac{3}{2}}}(5 + 3\gamma^2(-20 + 7\bar{\alpha}_2 + 3\bar{\alpha}_3) - 3\gamma^4(-40 + 15\bar{\alpha}_2 + 2\bar{\alpha_3} ) + \gamma^6(-64 + 24 \bar{\alpha}_2 + 3 \bar{\alpha}_3)) \\
    \chi_4 &= \frac{3\pi}{128}(35 + \gamma^4(1155 - 784\bar{\alpha}_2 + 24\bar{\alpha}_2^2 - 160\bar{\alpha}_3 - 16\bar{\alpha}_4) - 2\gamma^2(315 - 248\bar{\alpha}_2 + 8\bar{\alpha}_2^2 - 64\bar{\alpha}_3 - 8\bar{\alpha}_4))
\end{align}

PS* Schwarzschild:
\begin{align}
    \chi_1 &= \frac{2\gamma^2 - 1}{\sqrt{\gamma^2 - 1}} \\
    \chi_2 &= -\frac{\pi}{8}(3 + \gamma^2(2\alpha_2 - 15)) \\
    \chi_3 &= \frac{1}{3(\gamma^2 - 1)^{\frac{3}{2}}}(5 + 3\gamma^2(-20 + 5\alpha_2 + \alpha_3) - 3\gamma^4(-40 + 11\alpha_2 + 2\alpha_3) + \gamma^6(-64 + 18\alpha_2 + 3\alpha_3)) \\
    \chi_4 &= \frac{3\pi}{128}(35 + \gamma^4(1155 - 504\alpha_2 + 24\alpha_2^2 - 112\alpha_3 - 16\alpha_4) + \gamma^2(-630 + 280\alpha_2 - 16\alpha_2^2 + 80\alpha_3 + 16\alpha_4))
\end{align}

Next, we report the values of the EOB deformation coefficients, after matching to the PM scattering angles.

PS, isotropic:
\begin{align}
    \bar{q}_2(\gamma,\nu) &= \frac{3}{2\Gamma}(5\gamma^2 - 1)\Big( 1 - \Gamma \Big) \\
    \bar{q}_3(\gamma, \nu) &= \frac{1}{6 \left(\gamma ^2-1\right)^2 \Gamma
   ^3}(-8 \gamma ^7 (31 \Gamma -45) \nu +6 \gamma ^5 (113 \Gamma -117) \nu +36 \gamma ^3
   (11-19 \Gamma ) \nu +  \\
   &\hspace{2em} 2 \gamma ^2 \Big( \Gamma  \left(74 \sqrt{\gamma ^2-1} \nu +198
   \nu -99\right) -198 \nu +99\Big) -32 \sqrt{\gamma ^2-1} \Gamma  \nu \notag \\
   &\hspace{2em} +20 \gamma ^6
   \left(\Gamma  \left(4 \sqrt{\gamma ^2-1} \nu +18 \nu -9\right)-18 \nu +9\right) -13
   \gamma ^4 \left(\Gamma  \left(16 \sqrt{\gamma ^2-1} \nu +54 \nu -27\right)-54 \nu
   +27\right) \notag \\
   &\hspace{2em}+ 12 \Big(-8 \gamma ^7+20 \gamma ^5-14 \gamma ^3+18 \sqrt{\gamma ^2-1}
   \gamma ^2+6 \sqrt{\gamma ^2-1}+8 \sqrt{\gamma ^2-1} \gamma ^6 \notag \\
   &\hspace{2em} -32 \sqrt{\gamma ^2-1}
   \gamma ^4+3 \gamma \Big) \Gamma  \nu  \cosh^{-1}(\gamma ) +2 \gamma  (127 \Gamma
   -27) \nu -54 \Gamma  \nu +27 \Gamma +54 \nu -27) \notag
\end{align}

PS, Schwarzschild:
\begin{align}
    q_2(\gamma, \nu) &= \frac{3 \left(5 \gamma ^2-1\right) (1 - \Gamma )}{2 \Gamma } \\
    q_3(\gamma, \nu) &= \frac{1}{6 \left(\gamma
   ^2-1\right)^2 \Gamma ^3}\Big( 4 \gamma ^7 (45-17 \Gamma ) \nu +6 \gamma ^5 (47 \Gamma -51) \nu -144 \gamma ^3 (3
   \Gamma -1) \nu \\
   & \hspace{2em} +4 \gamma ^2 \left(\Gamma  \left(37 \sqrt{\gamma ^2-1} \nu +36 \nu
   -18\right)-36 \nu +18\right)-32 \sqrt{\gamma ^2-1} \Gamma  \nu \notag \\
   &\hspace{2em} +10 \gamma ^6
   \left(\Gamma  \left(8 \sqrt{\gamma ^2-1} \nu +18 \nu -9\right)-18 \nu
   +9\right)+\gamma ^4 \Big(\Gamma  \Big(-208 \sqrt{\gamma ^2-1} \nu \notag \\
   & \hspace{2em} -306 \nu
   +153\Big)+153 (2 \nu -1)\Big)+12 \Big(-8 \gamma ^7+20 \gamma ^5-14 \gamma ^3+18
   \sqrt{\gamma ^2-1} \gamma ^2 \notag \\
   &\hspace{2em} +6 \sqrt{\gamma ^2-1}+8 \sqrt{\gamma ^2-1} \gamma ^6-32
   \sqrt{\gamma ^2-1} \gamma ^4+3 \gamma \Big) \Gamma  \nu  \cosh ^{-1}(\gamma ) \notag \\
   & \hspace{2em} +2
   \gamma  (109 \Gamma -9) \nu -18 \Gamma  \nu +9 \Gamma +18 \nu -9 \Big) \notag
\end{align}    

PS*, isotropic:
\begin{align}
    \bar{\alpha}_2(\gamma, \nu) &= \frac{3 \left(5 \gamma ^2-1\right) (1 - \Gamma)}{2 \gamma ^2 \Gamma } \\
    \bar{\alpha}_3(\gamma, \nu) &= \frac{1}{6 \gamma ^2
   \left(\gamma ^2-1\right)^2 \Gamma ^3}\Big(16 \gamma ^7 (45-38 \Gamma ) \nu +6 \gamma ^5 (245 \Gamma -249) \nu -36 \gamma ^3
   (33 \Gamma -25) \nu \\
   &\hspace{2em} +2 \gamma ^2 \left(\Gamma  \left(74 \sqrt{\gamma ^2-1} \nu +450
   \nu -225\right)-450 \nu +225\right)-32 \sqrt{\gamma ^2-1} \Gamma  \nu \notag \\
   &\hspace{2em} +40 \gamma ^6
   \left(\Gamma  \left(2 \left(\sqrt{\gamma ^2-1}+9\right) \nu -9\right)-18 \nu
   +9\right) \notag \\ 
   &\hspace{2em} +\gamma ^4 \left(\Gamma  \left(747-2 \left(104 \sqrt{\gamma ^2-1}+747\right)
   \nu \right)+747 (2 \nu -1)\right) \notag \\
   &\hspace{2em} +12 \Big(-8 \gamma ^7+20 \gamma ^5-14 \gamma ^3+18
   \sqrt{\gamma ^2-1} \gamma ^2+6 \sqrt{\gamma ^2-1} \notag \\ 
   &\hspace{2em} +8 \sqrt{\gamma ^2-1} \gamma ^6-32
   \sqrt{\gamma ^2-1} \gamma ^4+3 \gamma \Big) \Gamma  \nu  \cosh ^{-1}(\gamma )+2
   \gamma  (163 \Gamma -63) \nu -126 \Gamma  \nu +63 \Gamma +126 \nu -63 \Big) \notag
\end{align}

PS*, Schwarzschild
\begin{align}
    \alpha_2(\gamma, \nu) &= \frac{3 \left(5 \gamma ^2-1\right) (\Gamma - 1)}{2 \gamma ^2 \Gamma } \\
    \alpha_3(\gamma, \nu) &= \frac{1}{6 \gamma ^2 \left(\gamma ^2-1\right)^2 \Gamma ^3} \Big(\gamma ^7 (540-428 \Gamma ) \nu +6 \gamma ^5 (179 \Gamma -183) \nu -72 \gamma ^3
   (13 \Gamma -9) \nu \\
   &\hspace{2em} +4 \gamma ^2 \left(\Gamma  \left(37 \sqrt{\gamma ^2-1} \nu +162
   \nu -81\right)-162 \nu +81\right)-32 \sqrt{\gamma ^2-1} \Gamma  \nu \notag \\
   & \hspace{2em} +10 \gamma ^6
   \left(\Gamma  \left(8 \sqrt{\gamma ^2-1} \nu +54 \nu -27\right)-54 \nu
   +27\right)+\gamma ^4 \Big(\Gamma  \Big(549  -2 \Big(104 \sqrt{\gamma
   ^2-1}+549\Big) \nu \Big) \notag \\
   &\hspace{2em} +549 (2 \nu -1)\Big)+12 \Big(-8 \gamma ^7+20 \gamma
   ^5-14 \gamma ^3+18 \sqrt{\gamma ^2-1} \gamma ^2+6 \sqrt{\gamma ^2-1}+8 \sqrt{\gamma
   ^2-1} \gamma ^6 \notag \\ 
   &\hspace{2em} -32 \sqrt{\gamma ^2-1} \gamma ^4+3 \gamma \Big) \Gamma  \nu  \cosh
   ^{-1}(\gamma )+10 \gamma  (29 \Gamma -9) \nu -90 \Gamma  \nu +45 \Gamma +90 \nu
   -45\Big) \notag
\end{align}

\section{Spinning Deformation Coefficients}

As a check of our methods, we reproduce the deformation coefficients of the SEOB-PM model. In addition to this we also provide the coefficients $\alpha_c$ of the centrifugal radius based model used in comparisons against the SEOB gauge choices. We note the difference of a factor of $\delta$ between the odd-in-$\chi_-$ results of our expressions and reference \cite{Buonanno:2024vkx}. This is simply due to the fact that for them, the odd in spin $\alpha$ coefficients are coefficients of $\delta \chi_-$, whereas we have absorbed the $\delta$. It is interesting to note that the deformation coefficients between the centrifugal-Kerr based model and the SEOB-PM model agree, with differences coming only from the factorization of the A-potential. The coefficients up to 3PM are given below.

\begin{align}
    \alpha^{(2)}_{(0,0)} = \frac{3}{2}\left(5 - \frac{1}{\gamma^2} \right) &- \frac{2 \theta^{(2)}_{(0,0)}}{\pi \gamma^2} \\
    \alpha^{(2)}_{(1,0)} = -\frac{\theta^{(2)}_{(1,0)}}{2\gamma\sqrt{\gamma^2 - 1}} & \qquad \alpha^{(2)}_{(0,1)} = -\frac{\theta^{(2)}_{(0,1)}}{2\gamma\delta \sqrt{\gamma^2 - 1}}
\end{align}
\begin{align}
    \alpha^{(3)}_{(0,0)} = \frac{12(5-11\gamma^2+6\gamma^4)\theta^{(2)}_{(0,0)} + \pi (35 - 204\gamma^2 + 309\gamma^4 - 142\gamma^6 - 3(\gamma^2-1)^{\frac{3}{2}}\theta^{(3)}_{(0,0)})}{6\pi\gamma^2(\gamma^2-1)^2} \\
    \alpha^{(3)}_{(1,0)} = \frac{\pi(5\gamma^2-3)\theta^{(2)}_{(1,0)} - 2\sqrt{\gamma^2-1}\theta^{(3)}_{(1,0)}}{2\pi\gamma(\gamma^2-1)^{\frac{3}{2}}} \qquad \alpha^{(3)}_{(0,1)} = \frac{\pi(5\gamma^2-3)\theta^{(2)}_{(0,1)} - 2\sqrt{\gamma^2-1}\theta^{(3)}_{(0,1)}}{2\pi\gamma\delta(\gamma^2-1)^{\frac{3}{2}}}\\
    \alpha^{(3)}_{(2,0)} = \frac{2 - 6\gamma^2 + 4\gamma^4 - \sqrt{\gamma^2 -1} \theta^{(2)}_{(2,0)}}{2\gamma^2\sqrt{\gamma^2 - 1}} \qquad \alpha^{(3)}_{(0,2)} = -\frac{\theta^{(3)}_{(0,2)}}{2\gamma^2\sqrt{\gamma^2-1}} \qquad \alpha^{(3)}_{(1,1)} = -\frac{\theta^{(3)}_{(1,1)}}{2\gamma^2\delta\sqrt{\gamma^2-1}}
\end{align}

The 4PM coefficients may be found in the attached Ancillary File. As can the coefficients for the centrifugal-PM model.

\section{The test mass limit}

In Sec.~\ref{subsection:seob+weob} it was stated that the test-mass limit of the SEOB type radial momentum tends to the Schwarzschild radial momentum, whereas the $w_{\rm eob}$ type models tend to a PM expanded Schwarzschild radial momentum. Here we expand upon this point. 

Consider the radial momentum for the SEOB-PM model
\begin{align}
    p_r^2 &= p_{\infty}^2 - \frac{\pphi^2}{r^2} + w_{\rm SEOB-PM}(r,\pphi,\gamma,\nu) \\
    w_{\rm SEOB-PM} &= \gamma^2 + \frac{\pphi^2}{r^2} + \frac{\gamma^2 B(r)}{A(r,\gamma,\nu)} - \frac{\pphi^2 B(r)}{r^2} - (B(r) - 1)
\end{align}
The deformation coefficients present in the $A$-potential vanish as the symmetric mass ratio $\nu \rightarrow 0$. As the effective $w$-potential is defined nonperturbatively, this reduces to the radial momentum of a test mass in a Schwarzschild background.

On the other hand, the $w_{\rm eob}$ model has radial momentum in the following form,
\begin{align}
    p_r^2 = p_{\infty}^2 - \frac{\pphi^2}{r^2} + w(r,\pphi,\gamma,\nu) \\
    w(r,\pphi,\gamma,\nu) = \sum_{i\geq 2} \frac{w_i(\gamma, \pphi, \nu)}{r^i},
\end{align}
where the coefficients $w_i$ are related to the deformation coefficients by the relations of Eq.~\eqref{w-potential-coefs-ps-isotropic}. The PS-isotropic deformation coefficients vanish in the test-mass limit $\nu -> 0$, therefore the effective $w$-potential coefficients reduce to,
\begin{align}
    w_1(\gamma) &= 2(2\gamma^2 - 1), \\
    w_2(\gamma) &= \frac{1}{2}(-3 + 15\gamma^2), \\
    w_3(\gamma) &= -\frac{1}{2} + 9\gamma^2, \\
    w_4(\gamma) &= \frac{1}{16}(-1 + 129\gamma^2),
\end{align}
Therefore the radial momentum in the test limit agrees with the \textit{PM expanded} isotropic gauge Schwarzschild radial momentum, rather than the full geodesic limit. It is interesting that in isotropic coordinates taking the PM expansion actually \textit{improves} the agreement to NR for the case of isotropic coordinates, rather than worsening it. 

Despite this, the Schwarzschild-PS* gauge (SEOB-PM) model has overall the best agreement with the numerical simulations.   

\section{\texorpdfstring{$w$ potential models}{w potential models}} \label{appendix: w-potentials}

In previous studies \cite{Damour:2022ybd, Rettegno:2023ghr}, the main object of focus for including post-Minkowskian deformations into the EOB formalism has been through the PM expanded $w$-potentials. These are defined by taking the PM expansion of the impetus formula, Eq.~\eqref{impetus-formula}),
\begin{align}
    p_r^2 = p_{\infty}^2 - \frac{\pphi^2}{r^2} + \sum_{i=1}^{n}\frac{w_{i}(\gamma, \nu)}{r^i}. 
\end{align}
Then the $w$-potential coefficients are determined from the scattering angle, by taking the PM expansion of the integral (using the same prescription described above)
\begin{align}
    \chi + \pi = -2 \pphi \int_{r_{\rm min}}^{\infty} \frac{dr}{r^2}\ \frac{1}{\sqrt{p_{\infty}^2 - \frac{\pphi^2}{r^2} + \sum_{i=1}^{n}\frac{w_{i}}{r^i}}}.
\end{align}
$\chi$ is expanded in the following way,
\begin{align}
    \chi = \sum_{n=0}^{\infty} 2\frac{\chi_n}{\pphi^n}.
\end{align}
The first few coefficients are given by \cite{Swain:2024ngs}:
\begin{align}
    \chi_1 &= \frac{w_1}{2p_{\infty}} \qquad \chi_2 = \frac{\pi w_2}{4} \\
    \chi_3 &= p_\infty w_3 + \frac{1}{2p_{\infty}}w_1 w_2 - \frac{1}{24p_{\infty}^3}w_1^3 \\
    \chi_4 &= \frac{3\pi}{16}(w_2^2 + 2 w_1 w_3 + 2p_{\infty}^2 w_4) \\
    \chi_5 &= \frac{1}{160 p_{\infty}^5}w_1^5 - \frac{1}{12p_{\infty}^3}w_1^3 w_2 + \frac{w_1}{2p_{\infty}}(w_2^2 + w_1 w_3)  \\
    & \nonumber \qquad+ 2p_{\infty}(w_2 w_3 + w_1 w_4) + \frac{4p_{\infty}^3}{3}w_5 \\
    \chi_6 &= \frac{5\pi}{32}(w_2^3 + 6 w_1 w_2 w_3 + 3w_1^2 w_4  \\
    & \nonumber \qquad + 3p_{\infty}^2(w_3^2 + 2w_2 w_4 + 2 w_1 w_5) + 3p_\infty^4 w_6)
\end{align}

From these relations the $w$-potential coefficients can be determined by matching to the known PM scattering angles. The $w$-coefficients in terms of PM scattering angle coefficients $\theta^{(n)}_{(i,j)}$ are given by:

\begin{align}
	w^{(1)}_{(0,0)} = \sqrt{\gamma^2 - 1} \theta^{(1)}_{(0,0)} 
\end{align}
\begin{align}
	w^{(2)}_{(0,0)} = \frac{2\theta^{(2)}_{(0,0)}}{\pi} \qquad
	w^{(2)}_{(1,0)} = \frac{\theta^{2}_{(1,0)}}{\sqrt{\gamma^2 - 1}} \qquad
	w^{(2)}_{(0,1)} = \frac{\theta^{(2)}_{(0,1)}}{\delta \sqrt{\gamma^2 - 1}}
\end{align}
\begin{align}
	w^{(3)}_{(0,0)} = \frac{\pi(\gamma^2 - 1) (\theta^{(1)}_{(0,0)})^3  - (\gamma^2 - 1) 24\theta^{(1)}	_{(0,0)} \theta^{(2)}_{(0,0)} + 12\pi (\gamma^2 - 1)\theta^{(3)}_{(0,0)}  }{24\pi(\gamma^2 - 1)^{\frac{3}{2}}}
\end{align}
\begin{align}
	w^{(3)}_{(1,0)} = \frac{-\pi \theta^{(1)}_{(0,0)} \theta^{(2)}_{(1,0)} + 2\theta^{(3)}_{(1,0)} }{\pi 	(\gamma^2 - 1)} \qquad
	w^{(3)}_{(0,1)} = \frac{  -\pi \theta^{(1)}_{(0,0)} \theta^{(2)}_{(0,1)} + 2 \theta^{(3)}_{(0,1)}   }{ \pi \delta (\gamma^2 - 1)    }
\end{align}
\begin{align}
	w^{(3)}_{(2,0)} = \frac{\theta^{(3)}_{(2,0)}}{2\sqrt{\gamma^2 - 1}} \qquad
	w^{(3)}_{(0,2)} = \frac{\theta^{(3)}_{(0,2)}}{2\sqrt{\gamma^2 - 1}} \qquad
	w^{(3)}_{(1,1)} = \frac{\theta^{(3)}_{(1,1)}}{2\delta \sqrt{\gamma^2 - 1}}
\end{align}
\begin{align}
	w^{(4)}_{(0,0)} = \frac{ -\pi^2 (\theta^{(1)}_{(0,0)})^4 + 24\pi (\theta^{(1)}_{(0,0)})^2 \theta^{(2)}_{(0,0)} 	- 48 (\theta^{(2)}_{(0,0)})^2 - 12\pi^2 \theta^{(1)}_{(0,0)} \theta^{(3)}_{(0,0)} + 32\pi \theta^{(4)}	_{(0,0)}   }{ 24\pi^2 (\gamma^2 - 1)  }
\end{align}
\begin{align}
	w^{(4)}_{(1,0)} &= \frac{ 9\pi (\theta^{(1)}_{(0,0)})^2 \theta^{(2)}_{(1,0)} - 24 \theta^{(2)}_{(0,0)}	\theta^{(2)}_{(1,0)} - 24\theta^{(1)}_{(0,0)}\theta^{(3)}_{(1,0)} + 4\pi \theta^{(4)}_{(1,0)} }	{ 8\pi(\gamma^2 - 1)^{\frac{3}{2}} } \\
	w^{(4)}_{(0,1)} &= \frac{9\pi (\theta^{(1)}_{(0,0)})^2 \theta^{(2)}_{(0,1)} - 24 \theta^{(2)}_{(0,0)} 	\theta^{(2)}_{(0,1)} - 24 \theta^{(1)}_{(0,0)}\theta^{(3)}_{(0,1)} + 4 \pi \theta^{(4)}_{(0,1)}   }{8\pi\delta (\gamma^2 - 1)^{\frac{3}{2}}} \\
    w^{(4)}_{(2,0)} &= \frac{-9\pi \left(\theta^{(2)}_{(1,0)}\right)^2 - 6\pi \theta^{(1)}_{(0,0)}\theta^{(3)}_{(2,0)} + 16\theta^{(4)}_{(2,0)}}{12\pi (\gamma^2 - 1)} \\
    w^{(4)}_{(1,1)} &= \frac{-9\pi \theta^{(2)}_{(0,1)}\theta^{(2)}_{(1,0)} - 3\pi \theta^{(1)}_{(0,0)}\theta^{(3)}_{(1,1)} + 8 \theta^{(4)}_{(1,1)}}{6\pi \delta (\gamma^2 - 1)} \\
    w^{(4)}_{(0,2)} &= \frac{-9\pi \left(\theta^{(2)}_{(0,1)}\right)^2 - 6\pi \theta^{(1)}_{(0,0) \theta^{(3)}_{(0,2)} + 16 \theta^{(4)}_{(0,2)}}}{12\pi (\gamma^2 - 1)}
\end{align}
\begin{align}
    w^{(4)}_{(3,0)} = \frac{\theta^{(4)}_{(3,0)}}{2(\gamma^2 - 1)^{\frac{3}{2}}} \qquad & w^{(4)}_{(0,3)} = \frac{\theta^{(4)}_{(0,3)}}{2\delta(\gamma^2 - 1)^{\frac{3}{2}}} \\
    w^{(4)}_{(1,2)} = \frac{\theta^{(4)}_{(1,2)}}{2\delta(\gamma^2 - 1)^{\frac{3}{2}}} \qquad &w^{(4)}_{(2,1)} = \frac{\theta^{(4)}_{(2,1)}}{2(\gamma^2 - 1)^{\frac{3}{2}}} 
\end{align}

We also discussed in the main text the possibility of deriving a $w_{\rm eob}$ type model by PM expanding an SEOB-PM type model. This would lead to a relationship between the coefficients $w^{(n)}_{(i,j)}$ and $\alpha^{(n)}_{(i,j)}$. We report here the mapping for the SEOB-PM $\alpha$ coefficients. 
\begin{align}
    w^{(1)}_{(0,0)} &= 2(2\gamma^2 - 1)\\
    w^{(2)}_{(0,0)} = -4 + 12\gamma^2 - \gamma^2 \alpha^{(2)}_{(0,0)} \qquad
    w^{(2)}_{(1,0)} &= -2\gamma \alpha^{(2)}_{(1,0)} \qquad 
    w^{(2)}_{(0,1)} = 2\gamma \alpha^{(2)}_{(0,1)} 
\end{align}
\begin{align}
    w^{(3)}_{(0,0)} = -8 - 2\pphi^2 + 32 \gamma^2 - \gamma^2 \alpha^{(3)}_{(0,0)} &\qquad
    w^{(3)}_{(1,0)} = -8\gamma \alpha^{(2)}_{(1,0)} + 2\gamma \alpha^{(3)}_{(1,0)} \\
    w^{(3)}_{(0,1)} = -8\gamma\alpha^{(2)}_{(0,1)} + 2\gamma \alpha^{(3)}_{(0,1)} &\qquad
    w^{(3)}_{(2,0)} = 4-6\gamma^2-\gamma^2\alpha^{(3)}_{(2,0)} \\
    w^{(3)}_{(0,2)} = \gamma^2 \alpha^{(3)}_{(0,2)} &\qquad
    w^{(3)}_{(1,1)} = -\gamma^2\delta\alpha^{(3)}_{(1,1)}
\end{align}
\begin{align}
    w^{(4)}_{(0,0)} &= -16 + 4\pphi^2 + 80\gamma^2 - 24\gamma^2\alpha^{(2)}_{(0,0)} +\gamma^2(\alpha^{(2)}_{(0,0)})^2 - 6\gamma^2\alpha^{(3)}_{(0,0)} - \gamma^2 \alpha^{(4)}_{(0,0)} \\
    w^{(4)}_{(1,0)} &= -2(12\gamma \alpha^{(2)}_{(1,0)} - \gamma\alpha^{(2)}_{(0,0)}\alpha^{(2)}_{(1,0)} + 4\gamma\alpha^{(3)}_{(1,0)} + \gamma\alpha^{(4)}_{(1,0)}) \\
    w^{(4)}_{(0,1)} &= -2(12\gamma\alpha^{(2)}_{(0,1)} - \gamma \alpha^{(2)}_{(0,0)}\alpha^{(2)}_{(0,1)} + 4\gamma\alpha^{(3)}_{(0,1)} + \gamma\alpha^{(4)}_{(0,1)})
    \end{align}
    \begin{align}
    w^{(4)}_{(2,0)} = 12 + 2\pphi^2 - 28\gamma^2 + 2\gamma^2\alpha^{(2)}_{(0,0)} - 6\gamma^2\alpha^{(3)}_{(2,0)} &\qquad
    w^{(4)}_{(0,2)} = -6\gamma^2 \alpha^{(3)}_{(0,2)} - \gamma^2 \alpha^{(4)}_{(0,2)}  \\
    w^{(4)}_{(1,1)} = -6\gamma^2 \alpha^{(3)}_{(1,1)} - \gamma^2 \alpha^{(4)}_{(1,1)} &\qquad
    w^{(4)}_{(3,0)} = 4\gamma \alpha^{(2)}_{(1,0)} - 2\gamma \alpha^{(4)}_{(3,0)} \\
    w^{(4)}_{(0,3)} = -2\gamma \alpha^{(4)}_{(0,3)} &\qquad
    w^{(4)}_{(2,1)} = 4\gamma\alpha^{(2)}_{(0,1)} + \gamma^{(4)}_{(2,1)} \\
    w^{(4)}_{(1,2)} = -2\gamma\alpha^{(4)}_{(1,2)}
\end{align}

\clearpage

\section{Additional Comparisons}
\label{appendix-additional-plots}

\begin{figure*}[htbp]
    \centering
    \includegraphics[width=\linewidth]{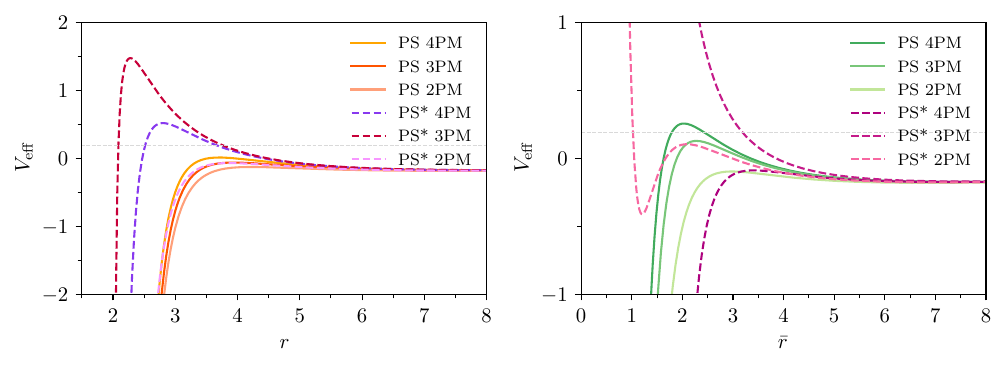}
    \caption{A comparison of the effective potentials, between PS and PS*, at the lower energy $\Gamma_1$, with angular momentum $\pphi \approx 4.5$, with equal mass. The left plot shows the Schwarzschild coordinate model, and the right the isotropic radial coordinate.}
    \label{fig:figure-8}
\end{figure*}

\begin{figure*}[htbp]
    \centering
    \includegraphics[width=\linewidth]{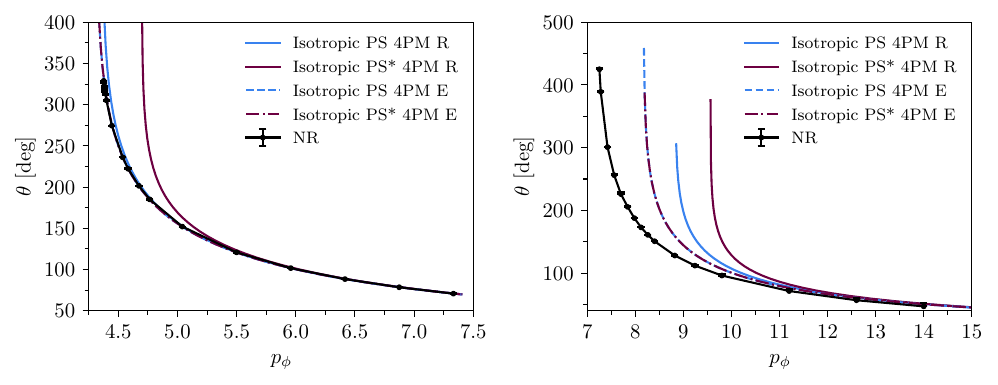}
    \caption{Comparing the scattering angles in isotropic coordinates between PS and PS* gauges, with the left plot showing the comparison at energy $\Gamma_1$ and the right $\Gamma_7$.
    \label{fig:PSvsStar-isotropic}}
\end{figure*}

\begin{figure*}[htbp]
    \centering
    \includegraphics[width=\linewidth]{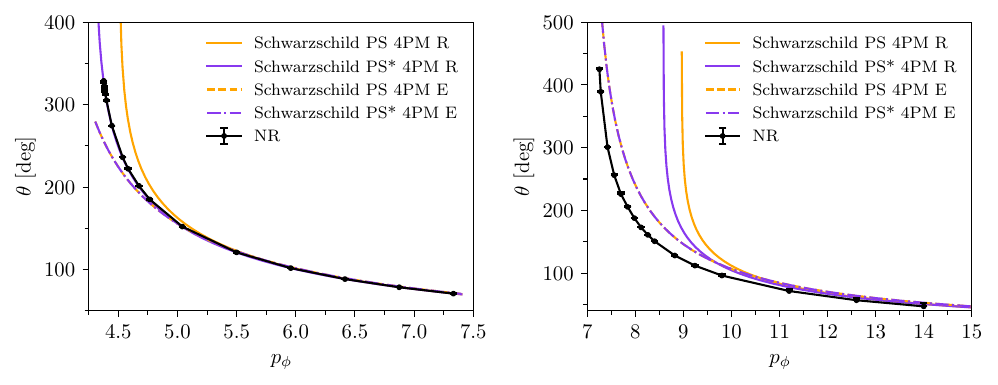}
    \caption{As above, but for the Schwarzschild coordinate gauge.
    \label{fig:PSvsStar-schwarzschild}}
\end{figure*}

\begin{figure*}[htbp]
    \centering
    \includegraphics[width=\linewidth]{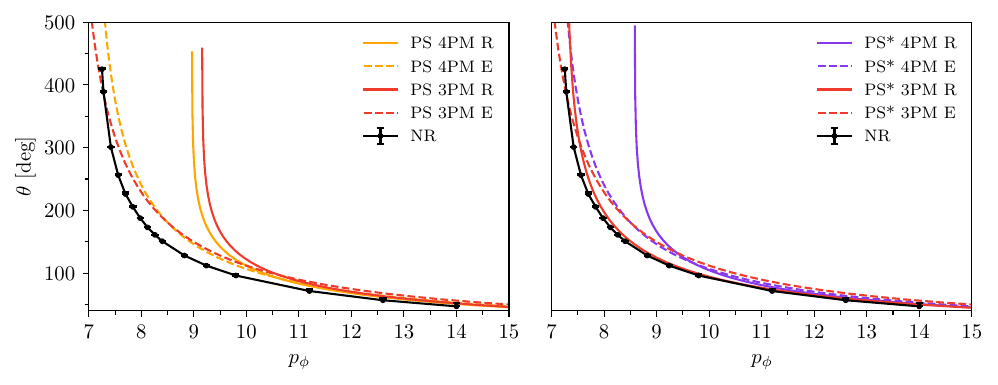}
    \caption{Schwarzschild coordinates, PS (left) vs PS* (right) at the higher energies. The solid lines represent the resummed $w$-potential, whereas the dashed denote the resummed model}
    \label{fig:schwarzschild_high_E_comparison}
\end{figure*}

\begin{figure}[htbp]
    \centering
    \includegraphics[width=\linewidth]{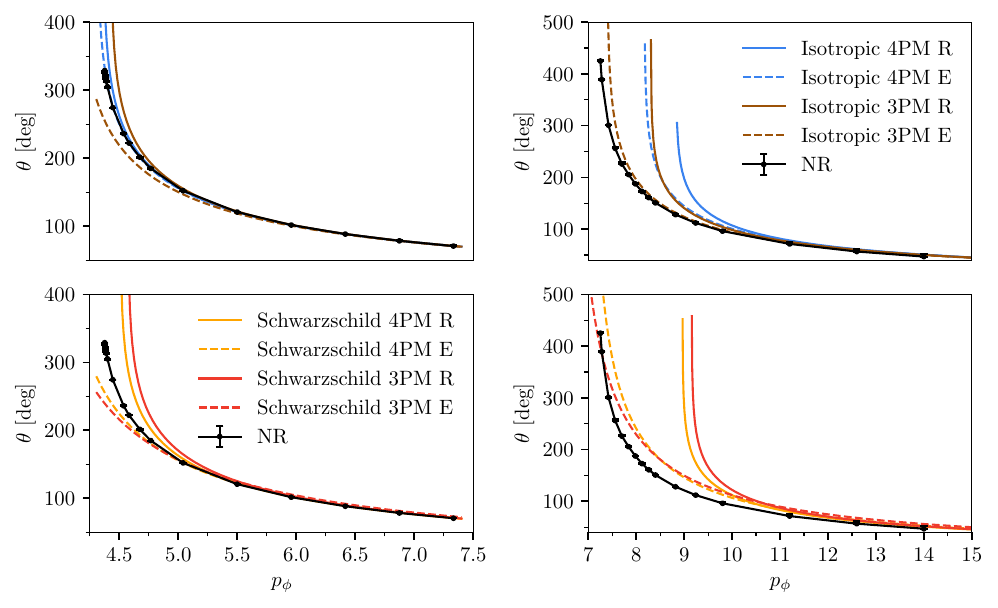}
    \caption{A comparison of the different PS gauge models. The left plots are at the lower energy $\Gamma_1$, and the right plots are the higher energy $\Gamma_7$. As usual, R denotes the use of a resummed model, whereas E denotes the PM expanded $w$-potential.}
    \label{fig:angle-comparison-PS-both-energies}
\end{figure}

\begin{figure}[htbp]
    \centering
    \includegraphics[width=\linewidth]{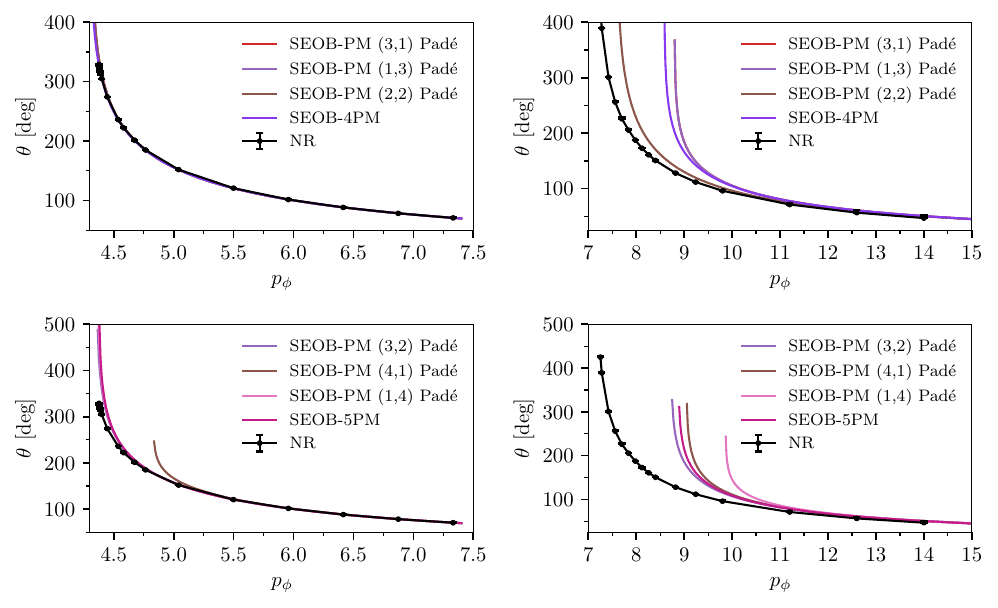}
    \caption{A comparison of the Pad\'e resummed SEOB models, across different PM orders and energies. The top plots are 4PM and the bottom are 5PM comparisons. Left shows the $\Gamma_1$ comparison, whereas the right shows $\Gamma_7$. As in the rest of the comparisons we take equal mass, $\nu = 0.25$.
    \label{fig:SEOB-Pade-comparisons}
    }
\end{figure}

\begin{figure}[htbp]
    \centering
    \includegraphics[width=\linewidth]{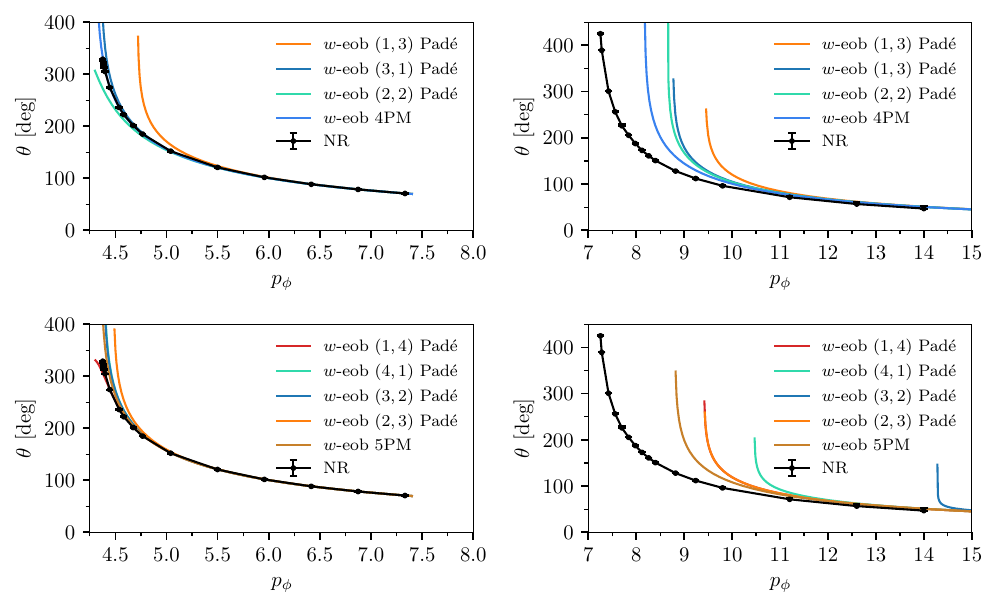}
    \caption{A comparison of the different Pad\'e resummations considered for the $w_{\rm eob}$ model, similarly to the previous plot we show 4PM at the top, 5PM on the bottom with left to right equating to low to high energies.
    \label{fig:weob-pade-comparisons}
    }
\end{figure}

\begin{figure}[htbp]
    \centering
    \includegraphics[width=0.5\linewidth]{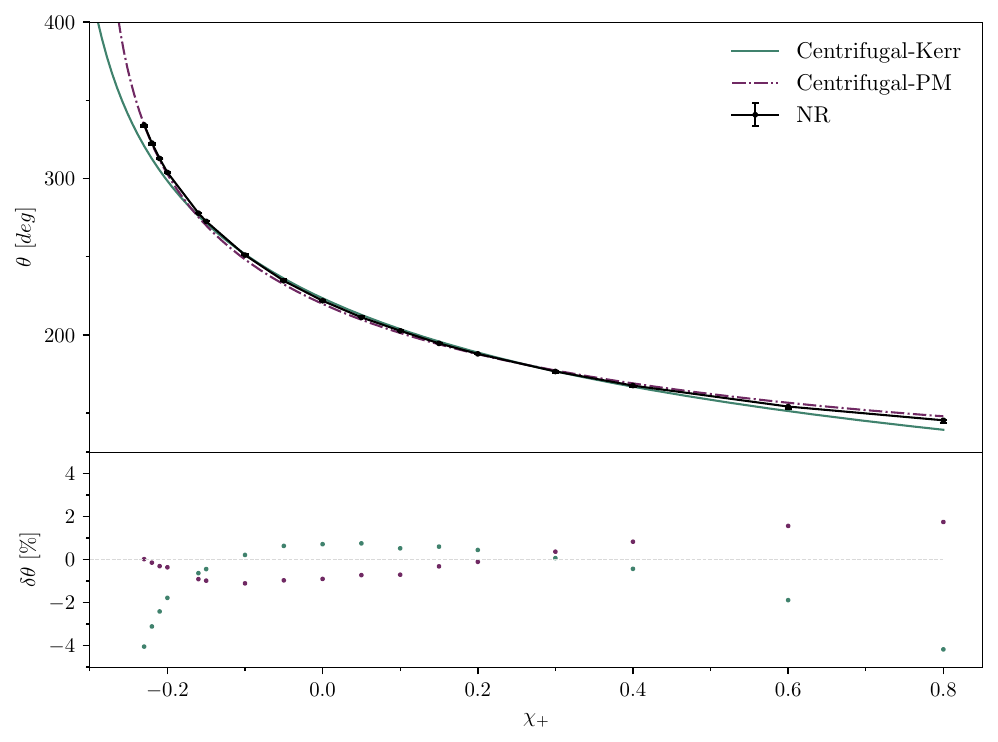}
    \caption{A comparison of the two centrifugal radius variants. We can see that the second variant (spin-deformed $r_c$) shows better agreement to NR.}
    \label{fig:centrifugal_variant_comparison}
\end{figure}

\begin{figure}[htbp]
   \centering
   \includegraphics[width=0.5\linewidth]{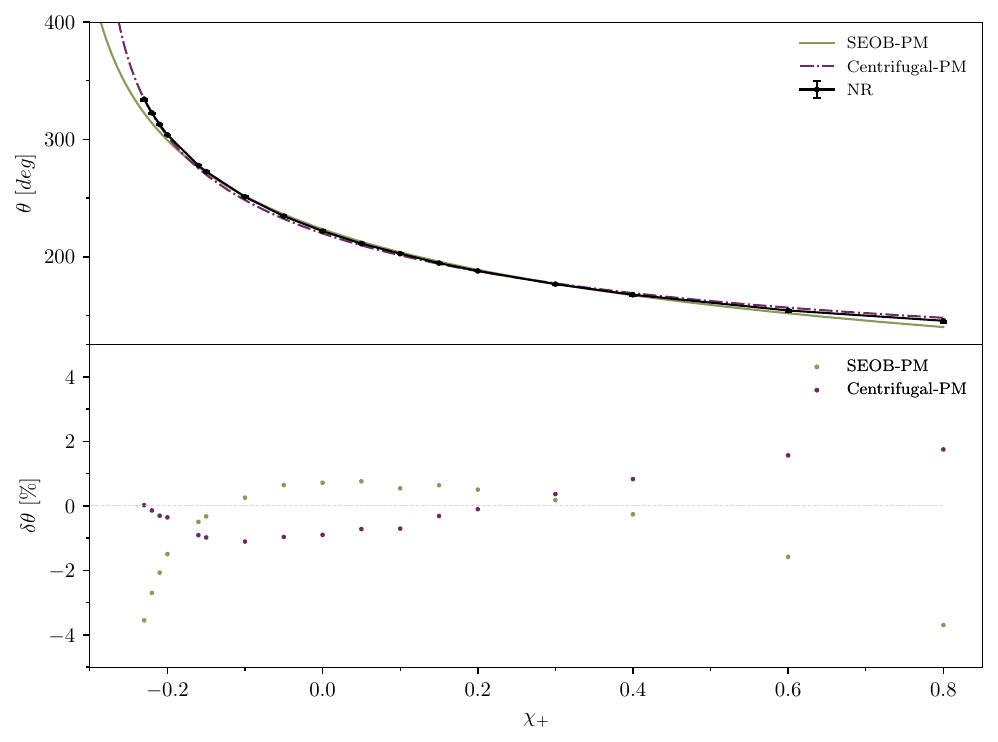}
   \caption{A comparison of \textit{centrifugal-PM} and SEOB-PM. The upper panel is the scattering angle with the lower panel giving the pointwise NR deviation as a percentage.
   \label{fig:spinning-NR-seob-centrifugal}}
\end{figure}

\begin{figure}[htbp]
    \centering
    \includegraphics[width=0.5\linewidth]{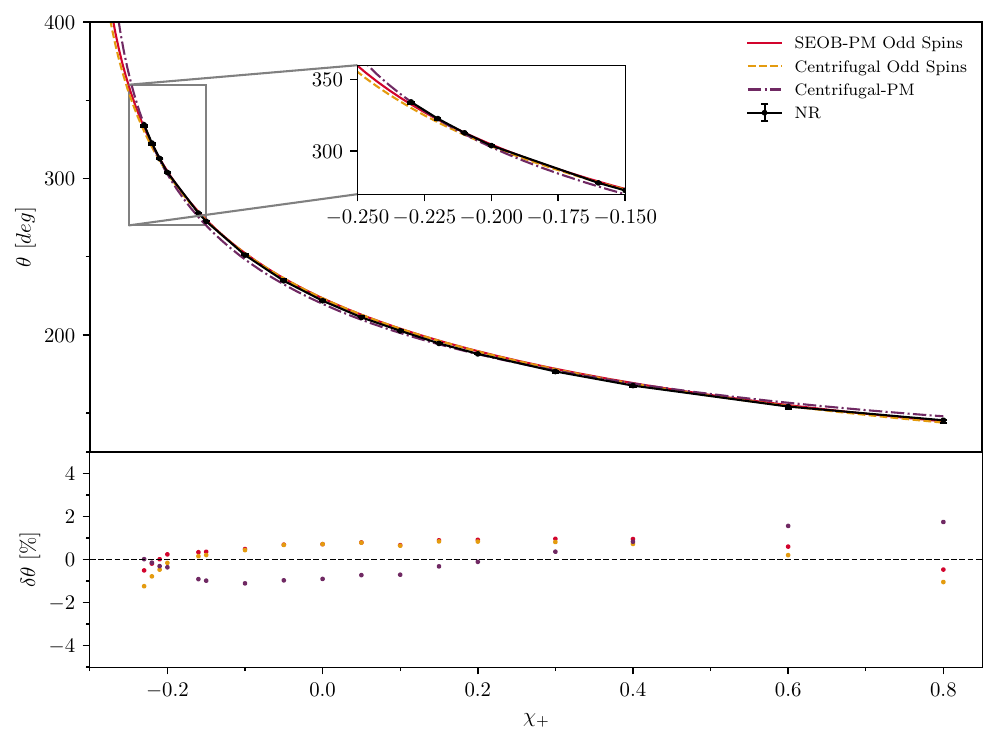}
    \caption{A comparison of the odd-spin variants of \textit{Centrifugal-Kerr} and SEOB-PM and the fully spinning \textit{Centrifugal-PM} model.
    \label{fig:spinning-NR-seob-odd-centrifugal}}
\end{figure}

\end{widetext}

\clearpage
\input{output.bbl}

\end{document}

%% file: output.bbl
%